\documentclass[11pt,final,onecolumn,twoside]{IEEEtran}
\usepackage{amsmath}
\usepackage{amssymb}
\usepackage{graphicx}
\usepackage{bm}
\usepackage{cite}
\newtheorem{theorem}{Theorem}[section]
\newtheorem{lemma}[theorem]{Lemma}

\newtheorem{assumption}         {Assumption}[section]

\begin{document}

\title{On Scaling Laws of Diversity Schemes in Decentralized Estimation}

\author{Alex S. Leong and Subhrakanti Dey
  \thanks{The authors are with the 
  Department of Electrical and Electronic Engineering,
  University of Melbourne, Parkville, Vic. 3010, Australia. E-mail {\tt \{asleong,sdey\}@unimelb.edu.au.} }
  \thanks{This work was supported by the Australian Research Council.}
  }
\maketitle
\thispagestyle{empty}
\begin{abstract}
This paper is concerned with decentralized estimation of a Gaussian source using multiple sensors. We consider a diversity scheme where only the sensor with the best channel sends their measurements over a fading channel to a fusion center, using the analog amplify and forwarding technique. The fusion centre reconstructs an 
MMSE estimate of the source based on the received measurements. A distributed version of the diversity scheme where sensors decide whether to transmit based  only on their local channel information is also considered. We derive asymptotic expressions for the expected distortion (of the MMSE estimate at the fusion centre) of these schemes as the number of sensors becomes large. For comparison, asymptotic expressions for the expected distortion for a coherent multi-access scheme and an orthogonal access scheme are  derived. We also study for the diversity schemes, the optimal power allocation for minimizing the expected distortion subject to average total power constraints. The effect of optimizing the probability of transmission on the expected distortion 
in the distributed scenario is also studied. It is seen that as opposed to the coherent 
multi-access scheme and the orthogonal scheme (where the expected distortion decays as $1/M$, $M$ being the number of sensors), the expected distortion 
decays only as $1/\ln(M)$ for the diversity schemes. This reduction of the decay rate can be seen as a tradeoff between the simplicity of the diversity 
schemes and the strict synchronization and large bandwidth requirements for the coherent multi-access and the orthogonal schemes, respectively. It is proved that optimal sensor transmit power allocation achieves the same asymptotic scaling law as the constant power allocation scheme, whereas it 
is observed that optimizing the sensor transmission probability (with or without optimal power allocation) in the distributed case makes
very little difference to the asymptotic scaling laws.   
\end{abstract}

\section{Introduction}
Wireless sensor networks have received much recent interest in the research community. Many different schemes for decentralized estimation of sources using multiple sensors have been proposed, e.g. \cite{RibeiroGiannakis1,SchizasGiannakisLuo,XiaoLuo,WuHuangLee,MergenTong,LiuElGamalSayeed}. One popular technique is analog amplify and forward \cite{GastparVetterli,Gastpar_JSAC}, where sensors transmit over fading channels a scaled version of their analog measurements to a fusion center, and has been shown to be optimal in some situations \cite{Gastpar_optimality}. Analog forwarding under different multiple access schemes such as coherent multi-access \cite{GastparVetterli,Xiao_coherent_TSP} and orthogonal access \cite{Cui_TSP}, with correlated data between sensors \cite{BahceciKhandani,FangLi}, and different network topologies \cite{ThatteMitra}, have also been studied. 

One problem with the analog amplify and forwarding technique is that it appears to be hard to implement, especially  when the number of sensors is large, e.g. it is difficult to synchronize a large number of sensors in the multi-access scheme (though studies suggest that even without perfect synchronization much of the gains can still be achieved \cite{Gastpar_JSAC,LiDai}), while there might not be a sufficiently large number of orthogonal channels available in the orthogonal scheme. This paper will study the performance of the analog forwarding technique using multiple access schemes which may be easier to implement, based on the concept of multi-user diversity \cite{KnoppHumblet1,ViswanathTseLaroia}.
Multi-user diversity refers to different users experiencing good channel conditions at different times, and can be exploited in the following manner: 
For the problem of maximizing the sum rate subject to average power constraints, the optimal solution is to schedule the users such that at most only one user transmits, with this user being the one having the best channel conditions at that instance. 
 
In this paper we will study the use of a similar diversity scheme in the decentralized estimation of a Gaussian source. In this scheme, which we will refer to as the \emph{multi-sensor diversity scheme}, the sensor with the best channel conditions at that time will amplify and forward its measurement to the fusion center, while the other sensors do not transmit. 
The multi-sensor diversity scheme requires knowledge of all the channel gains in order to decide on the best channel. A distributed version of the multi-sensor diversity scheme, similar to a distributed version of the multi-user diversity scheme studied in \cite{QinBerry} (see also \cite{AdireddyTong}) called the \emph{channel-aware ALOHA scheme}, will then also be considered. 

In this paper we are interested in the asymptotic behaviour of such schemes as the number of sensors $M$ goes to infinity. It is shown that in many cases the expected distortion (where the expectation is with respect to the time varying channel gains) decays to a non-zero limit at the rate $1/\ln(M)$. 
As a comparison we will also derive the expected distortion of the multi-access  and orthogonal access schemes, which decay at the rate $1/M$ for large $M$. These results are similar to the existing asymptotic results for the distortion in the multi-access scheme \cite{GastparVetterli,Gastpar_JSAC,LiuElGamalSayeed} and orthogonal access scheme \cite{Cui_TSP}, however the \emph{expected distortion} is not considered explicitly in these works. Note also that characterising performance via expectations has also been used in e.g. Kalman filtering with intermittent observations \cite{Sinopoli}, where the behaviour of the expected error covariance was studied. Another related concept is the distortion exponent \cite{Laneman}, which relates the expected distortion with SNR under different source and channel encodings, as the SNR goes to infinity.

We will also be interested in deriving the optimal power allocation to minimize the expected distortion subject to average power constraints. We will study this problem for the multi-sensor diversity and channel-aware ALOHA schemes. For the channel-aware ALOHA scheme, we will also consider the problem of optimizing the thresholds which determine when individual sensors will transmit. The effect of these optimal power allocation and/or optimal threshold selection schemes on the asymptotic scaling laws 
for the expected distortion will be studied in detail. It will be shown that with optimal power allocation, the asymptotic scaling law of expected distortion remains the 
same as that with constant power allocation policy.  It is also 
observed via numerical studies that with the optimal threshold selection in the distributed case (with or without optimal power allocation), the asymptotic scaling laws for expected distortion are very similar to that with an identical transmission probability of $1/M$ across all the sensors and some weaker theoretical results are proved. 

The paper is organised as follows. Section \ref{model_sec} specifies our model and the different multiple access schemes used by the sensors to communicate to the fusion center. Section \ref{asymptotic_sec} derives for symmetric parameters the asymptotic behaviour for the multi-sensor diversity, channel aware ALOHA, multi-access, and orthogonal access schemes, followed by comparisons and discussions. We comment on whether the results for the symmetric case can be extended to general parameters in Section \ref{general_sec}. Optimal power allocation for the multi-sensor and channel aware ALOHA schemes are considered in Section \ref{optim_sec}. It turns out that the performance of the simple constant power allocation of Section  \ref{asymptotic_sec} is very close that with the optimal power allocation, and we will prove why this is the case.  In Section \ref{threshold_sec} we study the problem of optimal threshold selection in the channel-aware ALOHA scheme and its effect on the asymptotic 
decay rate of the expected distortion. Finally, Section VII presents some concluding remarks and future research directions.

\section{System models}
\label{model_sec}
We wish to estimate a discrete time scalar signal $\theta_k$ modelled as an i.i.d. bandlimited Gaussian source  with zero mean and variance $\sigma_\theta^2$, with $k$ representing the time index. The Gaussian source is measured by $M$ sensors with sensor $i$ having measurements 
$$y_{i,k} = \theta_k + v_{i,k}, i=1, \dots, M$$ with $v_{i,k}$ being i.i.d. Gaussian with zero mean and noise variance $\sigma_i^2$, with $v_{i,k}$ independent of $v_{j,k}$ for $i\neq j$. Let $g_{i,k}$ be the randomly time-varying channel power gains from sensor $i$ to the fusion center, and $\alpha_{i,k}$ the amplification factors in the amplify and forward scheme. We assume that $g_{i,k}$ and $g_{j,k}$ are independent for $i\neq j$. 
The transmit power of sensor $i$ at time $k$ is defined as
$$\gamma_{i,k} = \alpha_{i,k}^2 \mathbb{E}[y_{i,k}^2] = \alpha_{i,k}^2 (\sigma_\theta^2+\sigma_i^2)$$
Next, we present the various multiple access schemes for transmitting the sensor measurements to a fusion center, considered in this paper.

\subsection{Multi-sensor diversity scheme}

Let $g_{max,k} = \max(g_{1,k},\dots, g_{M,k})$, and $i^*$ the index of the corresponding sensor. Consider a scheme where only the sensor with the best channel transmits its measurement to the fusion center.
The fusion center then receives
$$z_k = \sqrt{g_{max,k}} \alpha_{i^*,k} (\theta_k+v_{i^*,k}) + n_k$$
where $n_k$ is i.i.d. Gaussian with zero mean and variance $\sigma_n^2$.
Using the linear MMSE estimator \cite{Kay_estimation}, the  mean squared error or \emph{distortion} at time $k$ can be easily shown to be
$$D_k = \left( \frac{1}{\sigma_\theta^2} + \frac{g_{max,k} \alpha_{i^*,k}^2}{g_{max,k} \alpha_{i^*,k}^2 \sigma_{i^*}^2 + \sigma_n^2} \right)^{-1}$$

\subsection{Channel-aware ALOHA scheme}
\label{aloha_model_sec}
The multi-sensor diversity scheme requires knowledge of all the channel gains in order to determine the sensor with the best channel. In practice this could be achieved by having each sensor transmitting a pilot signal to the fusion center, which may then be used by the fusion center to estimate the individual channel gains. The fusion center can then determine and inform the sensor that it has the best channel. However, as the number of sensors increases, there is increasing overhead involved and the multi-sensor diversity scheme may be prohibitive for large networks, see e.g. \cite{QinBerry}. 

We consider now a scheme that we will call the channel-aware ALOHA scheme, that is based on a distributed scheme for multi-user diversity considered in \cite{QinBerry}, see also \cite{HongLeiChi} for a similar scheme in the distributed estimation of a constant parameter. In this scheme a sensor will forward its measurement to the fusion center only if 
$g_{i,k} > T_i$ for some threshold $T_i$. 

In \cite{QinBerry}, choosing $T_i$ such that $\Pr(g_{i,k} > T_i) = 1/M, \forall i$, was shown to be asymptotically optimal, in the sense that this gives the same rate of throughput scaling as in the multi-user diversity scheme, but with a fraction of throughput loss of $1/e$ (asymptotically). For much of this paper we will also use this choice of $T_i$. We discuss in Section \ref{threshold_sec} how the transmission threshold can be optimized and the effect of the optimal threshold on the 
expected distortion and its scaling law for large $M$ (assuming identical threshold for all sensors). 

In this scheme, if more than one sensor transmits, then a collision is assumed (whereby the fusion centre does not receive anything) and $D_k = \sigma_\theta^2$. Similarly if no sensor transmits then also $D_k = \sigma_\theta^2$. If only one sensor transmits, then 
$$D_k = \left( \frac{1}{\sigma_\theta^2} + \frac{g_{i^*,k} \alpha_{i^*,k}^2}{g_{i^*,k} \alpha_{i^*,k}^2 \sigma_{i^*}^2 + \sigma_n^2} \right)^{-1}$$
where $i^*$ is the index of the sensor that is transmitting. 

\subsection{Multi-access scheme}
In the (coherent) multi-access scheme \cite{GastparVetterli,Xiao_coherent_TSP}, the sensors transmit their measurements to the fusion center using the amplify and forward technique over a multi-access channel, so the fusion center receives the sum
\begin{equation}
\label{mac_coherent_sum}
 z_k = \sum_{i=1}^M \sqrt{g_{i,k}} \alpha_{i,k} (\theta_{k} + v_{i,k}) + n_k 
\end{equation}
The distortion at time $k$ is given by 
\begin{eqnarray*}
D_k & = & \left( \frac{1}{\sigma_\theta^2} + \frac{\left(\sum_{i=1}^M \sqrt{g_{i,k}} \alpha_{i,k} \right)^2}{\sum_{i=1}^M g_{i,k} \alpha_{i,k}^2 \sigma_i^2 + \sigma_n^2} \right)^{-1} \\
& = & \frac{\sigma_\theta^2 \left(\sum_{i=1}^M g_{i,k} \alpha_{i,k}^2 \sigma_i^2 + \sigma_n^2 \right)}{\sum_{i=1}^M g_{i,k} \alpha_{i,k}^2 \sigma_i^2 + \sigma_n^2 + \sigma_\theta^2 \left(\sum_{i=1}^M \sqrt{g_{i,k}} \alpha_{i,k} \right)^2}
\end{eqnarray*}

\subsection{Orthogonal access scheme}
In the orthogonal access scheme \cite{Cui_TSP}, the sensors transmit their measurements to the fusion center via orthogonal channels, so that the fusion center receives
$$z_{i,k} = \sqrt{g_{i,k}} \alpha_{i,k} (\theta_k + v_{i,k}) + n_{i,k}, i = 1, \dots, M$$
where $n_{i,k}$ is i.i.d. Gaussian with zero mean and variance $\sigma_n^2, \forall i$.
The distortion at time $k$ is given by 
$$D_k = \left( \frac{1}{\sigma_\theta^2} + \sum_{i=1}^M \frac{g_{i,k} \alpha_{i,k}^2}{g_{i,k} \alpha_{i,k}^2 \sigma_i^2 + \sigma_n^2}\right)^{-1}$$

\section{Asymptotic analysis}
\label{asymptotic_sec}
In this section, we are interested in deriving asymptotic expressions for 
$\mathbb{E}[D_k]$ as $M\rightarrow \infty$, where the expectation is over the random channel gains $g_{i,k}$, for the different schemes of Section \ref{model_sec}. Due to the i.i.d. (in time) nature of the models we will drop the subscript $k$. For analytical tractability we will first analyze ``symmetric'' sensor networks with $\sigma_{i}^2=\sigma_v^2, \forall i$, with the $g_i$'s being identically distributed, and simple power allocation policies, e.g. constant power allocation. See Section \ref{general_sec} for remarks on more general asymmetric situations, and Section \ref{optim_sec} for optimal power allocation. Apart from the multi-access scheme, for the other schemes we will need to assume a specific distribution in order to obtain precise asymptotic results. In these cases we will assume Rayleigh fading (i.e, the 
channel power gains are exponentially distributed), though most of our analytical methods can be adapted to other fading distributions. 

\emph{Notation}:
For two functions $f(t)$ and $g(t)$, we will use the standard asymptotic notation (see e.g. \cite{Olver}) and say that $f \sim g$ as $t \rightarrow t_0$, if $ \frac{f(t)}{g(t)} \rightarrow 1$ as $t \rightarrow t_0$. It is well known that the asymptotic relation $\sim$ is retained under addition,  multiplication and division. 

\emph{Notation}:
Extending the use of the symbol $\sim$ to functions of random variables, for functions $f(t,\omega)$ and $g(t,\omega)$, we will also say that $f \sim g \textrm{ w.p.1}$ as $t \rightarrow t_0$, if $\frac{f(t,\bullet)}{g(t,\bullet)} \rightarrow 1 \textrm{ w.p.1}$ as $t \rightarrow t_0$. For instance, if $X_i$ are i.i.d., then $\sum_{i=1}^M X_i \sim M \mathbb{E}[X_1] \textrm{ w.p.1}$ as $M \rightarrow \infty$, which follows from the definition and the strong law of large numbers. 

\subsection{Multi-sensor diversity scheme}
\label{diversity_sec}
Let us use $\alpha_{i^*}=1$ (constant power allocation), and $\alpha_j=0, \forall j \neq i^*$. Considering Rayleigh fading, we first have the following Lemma:
\begin{lemma}
\label{gmax_lemma}
Suppose the $g_i$'s are exponentially distributed with mean $1/\lambda$, and let $b>0$ be a constant. Then 
$$\mathbb{E} \left[\frac{1}{g_{max}+b}\right] \sim \frac{\lambda}{\lambda b+\ln{M}} \sim \frac{\lambda}{\ln(M)} \textrm{ as } M \rightarrow \infty$$
\end{lemma}
See Appendix \ref{gmax_lemma_appendix} for the proof of Lemma \ref{gmax_lemma}. 

\emph{Remark:} The expectation  above can actually be evaluated exactly as
\begin{equation}
\label{diversity_expectation_exact}
\mathbb{E} \left[\frac{1}{g_{max}+b}\right] = \sum_{k=0}^{M-1} M \lambda \binom{M-1}{k} (-1)^k \exp(\lambda(k+1)b) E_1(\lambda(k+1)b)
\end{equation}
where $E_1(.)$ is the exponential integral. 
However characterising the behaviour of $\mathbb{E} \left[\frac{1}{g_{max}+b}\right]$ as $M$ becomes large from the exact expression (\ref{diversity_expectation_exact}) does not appear obvious. 

With the help of Lemma \ref{gmax_lemma}, one can now prove the following result.
\begin{theorem}
Suppose the $g_i$'s are exponentially distributed with mean $1/\lambda$. Then in the multi-sensor diversity scheme with $\alpha_{i^*}=1$, and $\alpha_j=0, \forall j \neq i^*$,
\begin{equation}
\label{diversity_asympt}
\mathbb{E}[D] \sim \frac{\sigma_\theta^2 \sigma_v^2}{\sigma_\theta^2+\sigma_v^2} \left[1+ \frac{\sigma_n^2 \sigma_\theta^2}{\sigma_v^2(\sigma_\theta^2+\sigma_v^2)}  \frac{\lambda}{\ln(M) }\right] \textrm{ as } M \rightarrow \infty.
\end{equation}

\end{theorem}
\begin{proof}
We have
\begin{eqnarray*}
D & = & \left( \frac{1}{\sigma_\theta^2} + \frac{g_{max} }{g_{max}  \sigma_v^2 + \sigma_n^2} \right)^{-1} \\
 & = & \frac{\sigma_\theta^2 ( g_{max} \sigma_v^2 + \sigma_n^2)}{g_{max}(\sigma_v^2+\sigma_\theta^2)+\sigma_n^2} \\
 & = & \frac{\sigma_\theta^2 \sigma_v^2}{\sigma_\theta^2 + \sigma_v^2} \left(1 + \frac{\frac{\sigma_n^2 \sigma_\theta^2}{\sigma_v^2(\sigma_\theta^2+\sigma_v^2)}}{g_{max}+\frac{\sigma_n^2}{\sigma_\theta^2+\sigma_v^2}}  \right) 
\end{eqnarray*}
Therefore
$$\mathbb{E}[D] =  \frac{\sigma_\theta^2 \sigma_v^2}{\sigma_\theta^2 + \sigma_v^2} \mathbb{E} \left[ 1 + \frac{\frac{\sigma_n^2 \sigma_\theta^2}{\sigma_v^2(\sigma_\theta^2+\sigma_v^2)}}{g_{max}+\frac{\sigma_n^2}{\sigma_\theta^2+\sigma_v^2}} \right]$$

Using Lemma \ref{gmax_lemma}, we then have 
\begin{eqnarray*}
\mathbb{E}[D] & \sim &  \frac{\sigma_\theta^2 \sigma_v^2}{\sigma_\theta^2+\sigma_v^2} \left[1+ \frac{\sigma_n^2 \sigma_\theta^2}{\sigma_v^2(\sigma_\theta^2+\sigma_v^2)}  \frac{\lambda}{\ln(M) +\frac{\lambda \sigma_n^2}{\sigma_\theta^2+\sigma_v^2}}\right] \nonumber\\
& \sim & \frac{\sigma_\theta^2 \sigma_v^2}{\sigma_\theta^2+\sigma_v^2} \left[1+ \frac{\sigma_n^2 \sigma_\theta^2}{\sigma_v^2(\sigma_\theta^2+\sigma_v^2)}  \frac{\lambda}{\ln(M) }\right] 
\end{eqnarray*}
\end{proof}
Hence as $M \rightarrow \infty$, the expected distortion goes to $\frac{\sigma_\theta^2 \sigma_v^2}{\sigma_\theta^2+\sigma_v^2}$ at the rate $1/\ln(M)$. 

\subsection{Channel-aware ALOHA scheme}
\label{aloha_asymptotic_sec}
Recall that for this scheme, $T_i$ is chosen such that $\Pr(g_{i} > T_i) = 1/M$. Again, with constant power allocation, let us use $\alpha_i=1$ if sensor $i$ transmits. By the symmetry of the situation it is clear that $T_i = T, \forall i$.  Note that by the choice of $T$ each sensor has probability $1/M$ of transmitting to the fusion center (some of which will result in collision), so the long term total (across sensors) average power usage is the same as in the multi-sensor diversity scheme. 
Considering Rayleigh fading, we have the following result:
\begin{theorem}
Suppose the $g_i$'s are exponentially distributed with mean $1/\lambda$. Let $\alpha_i=1$ if sensor $i$ transmits. Then in the channel-aware ALOHA scheme
\begin{equation}
\label{aloha_asympt}
\mathbb{E}[D] \sim \sigma_\theta^2 (1-\frac{1}{e}) + \frac{1}{e} \frac{\sigma_\theta^2 \sigma_v^2}{\sigma_\theta^2+\sigma_v^2} \left[1 + \frac{\sigma_n^2 \sigma_\theta^2}{\sigma_v^2(\sigma_\theta^2+\sigma_v^2)} \frac{\lambda}{\ln M } \right]
\textrm{ as } M \rightarrow \infty. 
\end{equation}
\end{theorem}
\begin{proof}
We have
\begin{equation*}
\begin{split}
\Pr(\textrm{no sensor transmits}) & = (\Pr(g_i<T))^M \\ & = (1-\frac{1}{M})^M \\
\Pr(\textrm{successful transmission}) & = M \Pr(\textrm{sensor } i \textrm{ transmits successfully}) \\ & = M \Pr(g_i > T) \prod_{j\neq i} \Pr(g_j < T) \\ &= M \frac{1}{M} (1-\frac{1}{M})^{M-1} = (1-\frac{1}{M})^{M-1} \\
\Pr(\textrm{collision}) & = 1 - (1-\frac{1}{M})^M - (1-\frac{1}{M})^{M-1}
\end{split}
\end{equation*}
Then
\begin{equation*}
\begin{split}
\mathbb{E}[D] & =  \sigma_\theta^2 \Pr(\textrm{no sensor transmits}) + \sigma_\theta^2 \Pr(\textrm{collision}) \\ & \phantom{aaa} + \frac{1}{\Pr(g_i>T)} \int_{T}^\infty \left( \frac{1}{\sigma_\theta^2} + \frac{g_i}{g_i \sigma_v^2 + \sigma_n^2} \right)^{-1} p(g_i) dg_i \times \Pr(\textrm{successful transmission}) \\
& = \sigma_\theta^2 \left[1 - (1-\frac{1}{M})^{M-1} \right] + M (1-\frac{1}{M})^{M-1} \int_{T}^\infty \left( \frac{1}{\sigma_\theta^2} + \frac{g_i}{g_i \sigma_v^2 + \sigma_n^2} \right)^{-1} p(g_i) dg_i \\
& = \sigma_\theta^2 \left[1 - (1-\frac{1}{M})^{M-1} \right] + M (1-\frac{1}{M})^{M-1} \int_{T}^\infty \frac{\sigma_\theta^2 \sigma_v^2}{\sigma_\theta^2+\sigma_v^2} \left( 1 + \frac{\frac{\sigma_n^2 \sigma_\theta^2}{\sigma_v^2(\sigma_\theta^2+\sigma_v^2)}}{g_i+\frac{\sigma_n^2}{\sigma_\theta^2+\sigma_v^2}} \right) p(g_i) dg_i
\end{split}
\end{equation*}
Since the $g_i$'s are exponentially distributed with mean $1/\lambda$, $T = \frac{1}{\lambda} \ln M$ and 
$$\int_{T}^\infty \frac{1}{g+b} \lambda \exp(-\lambda g) dg  = \lambda \exp(\lambda b) E_1(\lambda(b+T)) = \lambda \exp(\lambda b) E_1(\lambda b+\ln M)$$
Hence 
\begin{eqnarray*}
\mathbb{E}[D] & = & \sigma_\theta^2 \left[1 - (1-\frac{1}{M})^{M-1} \right] \nonumber \\ & \phantom{aaa} & + M (1-\frac{1}{M})^{M-1}  \left[\frac{\sigma_\theta^2 \sigma_v^2}{\sigma_\theta^2+\sigma_v^2}\frac{1}{M} + \frac{\sigma_n^2 \sigma_\theta^4}{(\sigma_\theta^2+\sigma_v^2)^2} \lambda \exp\left(\frac{\lambda\sigma_n^2}{\sigma_\theta^2+\sigma_v^2}\right) E_1 \left(\frac{\lambda \sigma_n^2}{\sigma_\theta^2+\sigma_v^2} + \ln M  \right)\right] \nonumber \\
& \sim & \sigma_\theta^2 (1-\frac{1}{e}) + \frac{1}{e} \frac{\sigma_\theta^2 \sigma_v^2}{\sigma_\theta^2+\sigma_v^2} \left[1 + \frac{\sigma_n^2 \sigma_\theta^2}{\sigma_v^2(\sigma_\theta^2+\sigma_v^2)} \frac{\lambda}{\ln M + \frac{\lambda \sigma_n^2}{\sigma_\theta^2+\sigma_v^2}} \right] \nonumber\\
& \sim & \sigma_\theta^2 (1-\frac{1}{e}) + \frac{1}{e} \frac{\sigma_\theta^2 \sigma_v^2}{\sigma_\theta^2+\sigma_v^2} \left[1 + \frac{\sigma_n^2 \sigma_\theta^2}{\sigma_v^2(\sigma_\theta^2+\sigma_v^2)} \frac{\lambda}{\ln M } \right] 
\end{eqnarray*}
as $M \rightarrow \infty$, where  we have used the asymptotic expansion 
$$ E_1(z) \sim \frac{e^{-z}}{z} \left( 1 - \frac{1}{z} + \frac{2}{z^2} -+ \dots \right).$$
\end{proof}
The expected distortion in this case goes to $\sigma_\theta^2 (1-\frac{1}{e}) + \frac{1}{e} \frac{\sigma_\theta^2 \sigma_v^2}{\sigma_\theta^2+\sigma_v^2}$ at the rate $1/\ln(M)$ as $M \rightarrow \infty$. 

\subsection{Multi-access scheme}
\label{mac_sec}
For fairness of comparison, let us use here the scaling  $\alpha_i=1/\sqrt{M},\forall i$, which will result in the same total long term average transmit power usage as the multi-sensor diversity and channel-aware ALOHA schemes. 
\begin{theorem}
Let $\alpha_i=1/\sqrt{M},\forall i$. Then in the multi-access scheme, 
\begin{equation}
\label{mac_asympt} 
\mathbb{E}[D] \sim \frac{\sigma_v^2 \mathbb{E} [g_1] + \sigma_n^2}{M (\mathbb{E}[\sqrt{g_1}])^2 }
\textrm{ as } M \rightarrow \infty. 
\end{equation}
\end{theorem} 
\begin{proof}
We have
\begin{eqnarray*}
D & = & \frac{\sigma_\theta^2 \left( \sigma_v^2 \frac{1}{M} \sum_{i=1}^M g_i + \sigma_n^2 \right)}{\sigma_v^2 \frac{1}{M} \sum_{i=1}^M g_i + \sigma_n^2 + M \sigma_\theta^2 \left(\frac{1}{M} \sum_{i=1}^M \sqrt{g_i} \right)^2} \\
& \sim & \frac{\sigma_\theta^2 (\sigma_v^2 \mathbb{E}[g_1] + \sigma_n^2) }{ \sigma_v^2 \mathbb{E} [g_1] + \sigma_n^2 + M \sigma_\theta^2 (\mathbb{E}[\sqrt{g_1}])^2} \textrm{ w.p.1 } 
\end{eqnarray*}
provided the expectations $\mathbb{E}[g_1]$ and $\mathbb{E}[\sqrt{g_1}]$ exist, where the last line comes from applying the strong law of large numbers and the definition and properties of $\sim$. 
Since $D$ is always bounded, we can then use results on uniform integrability, e.g. \cite{GrimmettStirzaker}, to conclude that 
\begin{eqnarray*}
\mathbb{E}[D] & \sim &  \frac{\sigma_\theta^2 (\sigma_v^2 \mathbb{E}[g_1] + \sigma_n^2) }{ \sigma_v^2 \mathbb{E} [g_1] + \sigma_n^2 + M \sigma_\theta^2 (\mathbb{E}[\sqrt{g_1}])^2} \nonumber \\
& \sim & \frac{\sigma_v^2 \mathbb{E} [g_1] + \sigma_n^2}{M (\mathbb{E}[\sqrt{g_1}])^2 } 
\end{eqnarray*}
\end{proof}
Thus the expected distortion decays to zero at the rate $1/M$ as $M \rightarrow \infty$, similar to the scaling behaviour for the distortion derived in \cite{Gastpar_JSAC}. 

\subsection{Orthogonal access scheme}
\label{orth_sec}
\begin{theorem}
Suppose the $g_i$'s are exponentially distributed with mean $1/\lambda$. Let $\alpha_i=1/\sqrt{M},\forall i$.  Then in the orthogonal access scheme
\begin{equation}
\label{orth_asympt}
\mathbb{E}[D] \sim \frac{1}{\frac{1}{\sigma_\theta^2} +\frac{1}{\lambda \sigma_n^2}} + \frac{2 \sigma_v^2}{M \lambda^2 \sigma_n^4 } \frac{1}{(\frac{1}{\sigma_\theta^2} +\frac{1}{\lambda \sigma_n^2})^2}
\textrm{ as } M \rightarrow \infty. 
\end{equation}
\end{theorem}
\begin{proof}
We have
\begin{eqnarray*}
D & = & \frac{1}{\frac{1}{\sigma_\theta^2} + \sum_{i=1}^M \frac{g_{i}/M} {g_{i}  \sigma_v^2/M + \sigma_n^2}} \\
& = & \frac{1}{\frac{1}{\sigma_\theta^2} + \sum_{i=1}^M \frac{g_{i}} {g_{i}  \sigma_v^2 + M \sigma_n^2}} \\
& \sim & \frac{1}{\frac{1}{\sigma_\theta^2} +  M \mathbb{E}[ \frac{g_1}{g_1 \sigma_v^2 + M \sigma_n^2} ]} \textrm{ w.p.1 }
\end{eqnarray*}
provided the expectation $\mathbb{E}[g_1/(g_1 \sigma_v^2 + \sigma_n^2)]$ (and hence $\mathbb{E}[g_1/(g_1 \sigma_v^2 + M\sigma_n^2)]$) exists, where the last line now comes from using a strong law of large numbers for triangular arrays \cite{HuMoriczTaylor}, and the definition and properties of $\sim$. 
Hence by uniform integrability
$$\mathbb{E}[D] \sim \frac{1}{\frac{1}{\sigma_\theta^2} +  M \mathbb{E}[ \frac{g_1}{g_1 \sigma_v^2 + M \sigma_n^2} ]}$$
If we now assume all $g_i$ are exponentially distributed with mean $1/\lambda$, then 
\begin{equation*}
\begin{split}
\mathbb{E}&\left[ \frac{g_1}{g_1 \sigma_v^2 + M \sigma_n^2} \right]  =  \mathbb{E} \left[\frac{1}{\sigma_v^2} \left( 1-\frac{M \sigma_n^2/\sigma_v^2}{g_1+M\sigma_n^2/\sigma_v^2} \right)   \right] \\
& =  \frac{1}{\sigma_v^2} \left[1 - \frac{\lambda M \sigma_n^2}{\sigma_v^2}\exp\left(\frac{\lambda M \sigma_n^2}{\sigma_v^2} \right) E_1 \left(\frac{\lambda M \sigma_n^2}{\sigma_v^2} \right)\right] \\ 
& \sim  \frac{1}{\sigma_v^2} \left(\frac{\sigma_v^2}{\lambda M \sigma_n^2} - \frac{2 \sigma_v^4}{\lambda^2 M^2 \sigma_n^4} \right)
\end{split}
\end{equation*}
Hence 
\begin{eqnarray*}
\mathbb{E}[D] & \sim & \frac{1}{\frac{1}{\sigma_\theta^2} +\frac{1}{\lambda \sigma_n^2} - \frac{2 \sigma_v^2}{\lambda^2 M \sigma_n^4}} \nonumber \\
& \sim & \frac{1}{\frac{1}{\sigma_\theta^2} +\frac{1}{\lambda \sigma_n^2}} + \frac{2 \sigma_v^2}{M \lambda^2 \sigma_n^4 } \frac{1}{(\frac{1}{\sigma_\theta^2} +\frac{1}{\lambda \sigma_n^2})^2}
\end{eqnarray*}
which converges to $(\frac{1}{\sigma_\theta^2} +\frac{1}{\lambda \sigma_n^2})^{-1}$ at the rate $1/M$. 
\end{proof}
The limit  $(\frac{1}{\sigma_\theta^2} +\frac{1}{\lambda \sigma_n^2})^{-1}$ for $D$ as $M \rightarrow \infty$ was also previously shown in \cite{Cui_TSP}, though the rate of convergence was not derived. 

\subsection{Comparisons and discussions}
The limit $\frac{\sigma_\theta^2 \sigma_v^2}{\sigma_\theta^2+\sigma_v^2} = (\frac{1}{\sigma_\theta^2} + \frac{1}{\sigma_v^2} )^{-1}$ in the multi-sensor diversity scheme corresponds to the distortion that can be achieved with a single sensor, with estimation performed at that sensor, i.e. no further analog forwarding to a fusion center. 
For the channel-aware ALOHA scheme, the limit is $\sigma_\theta^2 (1-\frac{1}{e}) + \frac{1}{e} \frac{\sigma_\theta^2 \sigma_v^2}{\sigma_\theta^2+\sigma_v^2}$, which is clearly larger than the limit in the multi-sensor diversity scheme. It can be regarded as a weighted combination of the limiting value $\frac{\sigma_\theta^2 \sigma_v^2}{\sigma_\theta^2+\sigma_v^2}$ when there is a successful transmission, and the distortion $\sigma_\theta^2$ when transmissions are unsuccessful, with $\frac{1}{e}$ being the asymptotic probability of successful transmission as $M \rightarrow \infty$ (which also corresponds to the asymptotic throughput of a slotted ALOHA system \cite{QinBerry,BertsekasGallager}). 
We note also that the limit $(\frac{1}{\sigma_\theta^2} +\frac{1}{\lambda \sigma_n^2})^{-1}$ in the orthogonal scheme using $\alpha_i = 1/\sqrt{M}, \forall i$ is different from the limit in the diversity scheme. Under the choices of $\alpha_i$ in this paper, the expected distortion goes to zero only in the multi-access scheme.\footnote{However if e.g. we use $\alpha_i=1, \forall i$, then the expected distortion will also go to zero in the orthogonal scheme.} 

In terms of speed of convergence, the rate $1/M$ is achieved in the multi-access and orthogonal schemes. On the other hand, we get a slower convergence rate of $1/\ln(M)$ in the diversity schemes. A similar  $1/\ln(M)$ rate is achieved when sensor measurements are transmitted to a fusion center digitally using separate source/channel coding, e.g. as in the CEO problem \cite{GastparVetterli,BergerZhangViswanathan}.

Finally, in regards to implementation, the multi-access scheme requires that we the measurements add up coherently as in (\ref{mac_coherent_sum}) at the fusion center, which may be difficult to achieve for large sensor networks, since it requires distributed transmit beamforming to be implemented. The orthogonal access scheme does not require as much synchronization between sensors \cite{Cui_TSP}, but each sensor will require its own orthogonal channel. The multi-sensor diversity scheme does not have these issues, though it will still require the fusion center to determine which sensor has the best channel, with this information then fed back to the sensors. The channel-aware ALOHA scheme is probably the easiest to implement in practice, however asymptotically it has larger expected distortion when compared to the multi-sensor diversity and multi-access schemes.  

\subsection{Numerical studies}
\label{numerical_sec}
Consider an example with $\sigma_\theta^2=1, \sigma_v^2=0.2, \sigma_n^2=0.1$, and let $g_i, \forall i$ be exponentially distributed with mean $1/2$. Note that then $\mathbb{E}[\sqrt{g_i}] = \sqrt{\pi/8}$, $(\frac{1}{\sigma_\theta^2} + \frac{1}{\sigma_v^2} )^{-1}=0.1667$, $(\frac{1}{\sigma_\theta^2} +\frac{1}{\lambda \sigma_n^2})^{-1}=0.1667$, and $\sigma_\theta^2 (1-\frac{1}{e}) + \frac{1}{e} \frac{\sigma_\theta^2\sigma_v^2}{\sigma_\theta^2+\sigma_v^2} = 0.6934$. 

\begin{figure}[tbp]
\centering
\includegraphics[width=8.0cm]{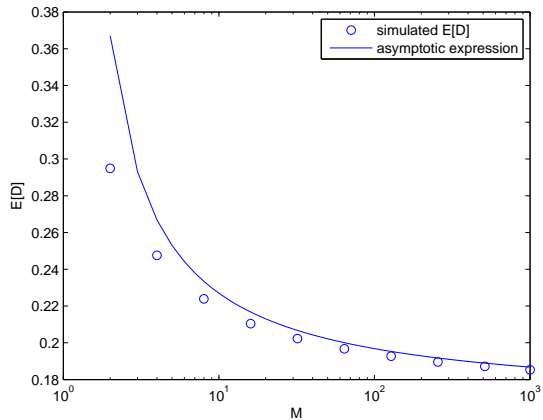}
\caption{Multi-sensor diversity scheme. Comparison between simulated expected distortion and asymptotic expression.}
\label{diversity_plot}
\end{figure}

\begin{figure}[tbp]
\centering
\includegraphics[width=8.0cm]{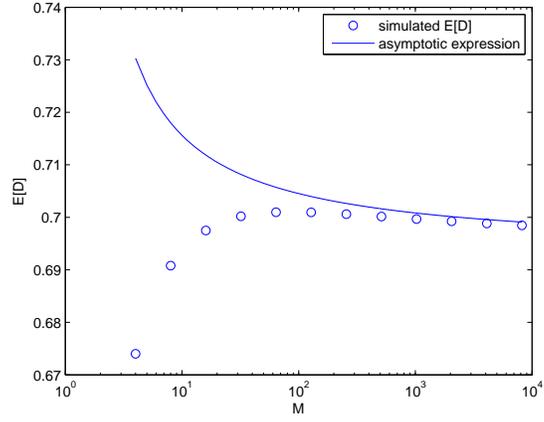}
\caption{Channel-aware ALOHA scheme. Comparison between simulated expected distortion and asymptotic expression.}
\label{aloha_plot}
\end{figure}

\begin{figure}[tbp]
\centering
\includegraphics[width=8.0cm]{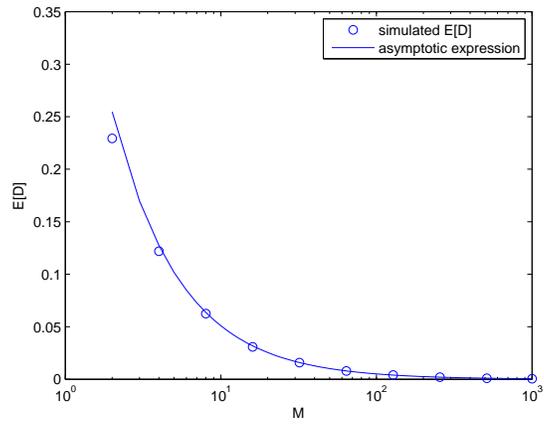}
\caption{Multi-access scheme. Comparison between simulated expected distortion and asymptotic expression.}
\label{mac_plot}
\end{figure}

\begin{figure}[tbp]
\centering
\includegraphics[width=8.0cm]{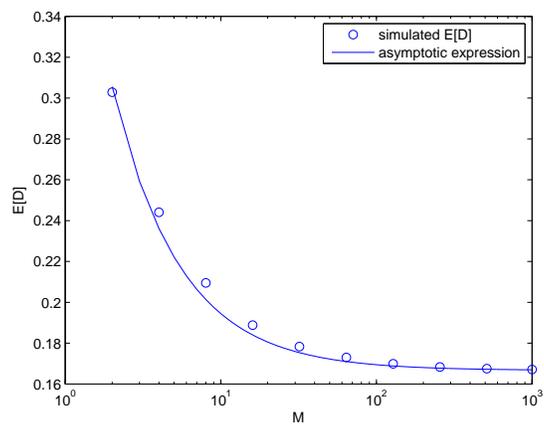}
\caption{Orthogonal access scheme. Comparison between simulated expected distortion and asymptotic expression.}
\label{orth_plot}
\end{figure}

In Fig. \ref{diversity_plot} we compare between the simulated expected distortion (averaging over 100000 channel realizations) and the asymptotic expression (\ref{diversity_asympt}) for the multi-sensor diversity scheme, for different numbers of sensors $M$. 
In Fig. \ref{aloha_plot} we compare between the simulated expected distortion and the asymptotic expression (\ref{aloha_asympt}) for the channel-aware ALOHA scheme, for different numbers of sensors $M$. 
In Fig. \ref{mac_plot} we compare between the simulated expected distortion and the asymptotic expression (\ref{mac_asympt}) for the multi-access scheme. 
 In Fig. \ref{orth_plot} we compare between the simulated expected distortion and the asymptotic expression (\ref{orth_asympt}) for the orthogonal access scheme. 
In each case, the validity of the respective asymptotic expressions for large $M$ is confirmed. 

We also see that in the channel-aware ALOHA scheme, the expected distortion is not necessarily monotonically decreasing with the number of sensors, though for large $M$ the $1/\ln(M)$ decay will still occur. 

\section{General parameters}
\label{general_sec}
In this section we investigate how the results of Section \ref{asymptotic_sec} change when the sensor noise variances are not necessarily identical, and the fading channels are not necessarily identically distributed.  The idea is to obtain upper and lower bounds on the expected distortion which asymptotically will have the same scaling behaviour, a similar method was used in \cite{LeongDeyEvans_TAES} in the context of linear state estimation.
We will see that in some cases the scaling behaviour derived in Section \ref{asymptotic_sec} is still preserved, while in other cases not much can be said in general.

\subsection{General sensor noise variances}
We consider here the case where the sensor noise variances $\sigma_i^2, i=1, \dots, M$ are not necessarily identical, though the fading channels are still assumed to be i.i.d. across sensors.  We assume that the sensor noise variances can be bounded from both above and below, i.e.
$$0 < \sigma_{min}^2 \leq \sigma_i^2 \leq \sigma_{max}^2 < \infty, \forall i$$
Then we note in all the different schemes considered here, $D$ is an increasing function of $\sigma_i^2$ for all $i$. Hence we can upper and lower bound $D$ with the symmetric results using $\sigma_i^2 = \sigma_{max}^2, \forall i$ and $\sigma_i^2 = \sigma_{min}^2, \forall i$ respectively. 

In the multi-sensor diversity scheme, we have for Rayleigh fading
\begin{equation*}
\begin{split}
\frac{\sigma_\theta^2 \sigma_{min}^2}{\sigma_\theta^2+\sigma_{min}^2} & \left[1+ \frac{\sigma_n^2 \sigma_\theta^2}{\sigma_{min}^2(\sigma_\theta^2+\sigma_{min}^2)}  \frac{\lambda}{\ln(M) }\right] (1+o(1)) \\ & \leq \mathbb{E}[D] \leq \frac{\sigma_\theta^2 \sigma_{max}^2}{\sigma_\theta^2+\sigma_{max}^2} \left[1+ \frac{\sigma_n^2 \sigma_\theta^2}{\sigma_{max}^2(\sigma_\theta^2+\sigma_{max}^2)}  \frac{\lambda}{\ln(M) }\right] (1+o(1)) 
\end{split}
\end{equation*}
Note that the upper and lower bounds do not converge to the same limit as $M \rightarrow \infty$, so for general sensor noise variances one can not say much more about its asymptotic behaviour. Indeed, one can construct situations that can be shown to not converge to any limit, in a similar fashion as in \cite{LeongDeyEvans_TAES}. 

In the channel-aware ALOHA scheme, for Rayleigh fading we have
\begin{equation*}
\begin{split}
\sigma_\theta^2(1-\frac{1}{e}) + \frac{1}{e} \frac{\sigma_\theta^2 \sigma_{min}^2}{\sigma_\theta^2+\sigma_{min}^2} & \left[1+ \frac{\sigma_n^2 \sigma_\theta^2}{\sigma_{min}^2(\sigma_\theta^2+\sigma_{min}^2)}  \frac{\lambda}{\ln(M) }\right] (1+o(1)) \\ & \leq \mathbb{E}[D] \leq \sigma_\theta^2(1-\frac{1}{e}) + \frac{1}{e} \frac{\sigma_\theta^2 \sigma_{max}^2}{\sigma_\theta^2+\sigma_{max}^2} \left[1+ \frac{\sigma_n^2 \sigma_\theta^2}{\sigma_{max}^2(\sigma_\theta^2+\sigma_{max}^2)}  \frac{\lambda}{\ln(M) }\right] (1+o(1)) 
\end{split}
\end{equation*}
Similarly, little more can be said about the asymptotic behaviour for the channel-aware ALOHA scheme in general. 

On the other hand, in the multi-access scheme, we will have
$$\frac{\sigma_{min}^2 \mathbb{E} [g_1] + \sigma_n^2}{M (\mathbb{E}[\sqrt{g_1}])^2 } (1+o(1)) \leq \mathbb{E}[D] \leq \frac{\sigma_{max}^2 \mathbb{E} [g_1] + \sigma_n^2}{M (\mathbb{E}[\sqrt{g_1}])^2 } (1+o(1))$$
Since the upper and lower bounds both converge to zero at the rate $1/M$, the general situation will also have the same scaling behaviour as the bounds. 

In the orthogonal scheme, we have for Rayleigh fading
$$\left(\frac{1}{\frac{1}{\sigma_\theta^2} +\frac{1}{\lambda \sigma_n^2}} + \frac{2 \sigma_{min}^2}{M \lambda^2 \sigma_n^4 } \frac{1}{(\frac{1}{\sigma_\theta^2} +\frac{1}{\lambda \sigma_n^2})^2} \right)(1+o(1)) \leq \mathbb{E}[D] \leq \left( \frac{1}{\frac{1}{\sigma_\theta^2} +\frac{1}{\lambda \sigma_n^2}} + \frac{2 \sigma_{max}^2}{M \lambda^2 \sigma_n^4 } \frac{1}{(\frac{1}{\sigma_\theta^2} +\frac{1}{\lambda \sigma_n^2})^2} \right) (1+o(1))$$
Here, both the upper and lower bounds converge to $(\frac{1}{\sigma_\theta^2} +\frac{1}{\lambda \sigma_n^2})^{-1}$ at the rate $1/M$, and so the general situation will also do so. 

\subsection{Non-identically distributed fading channels}
In the previous subsection where the sensor noise variances were different but the fading channels were still i.i.d., we found that the asymptotic behaviour was still preserved in the multi-access and orthogonal access schemes. We now consider the situation where the sensor noise variances are identical ($\sigma_i^2=\sigma_v^2, \forall i$), and the fading channels are independent but not necessarily identically distributed, though for tractability assuming that the fading distributions belong to the same ``family''. To be more specific, we make the following assumption:
\begin{assumption}
\label{fading_assumption}
The channel gains $g_i$ can be written as
$$ g_i = \mu_i h_i, \forall i,$$
where $\mu_i > 0$ are constants satisfying
$$0 < \mu_{min} \leq \mu_i \leq \mu_{max} < \infty,$$ and the $h_i$'s are identically distributed. 
\end{assumption}

For example, if $g_i$ is exponentially distributed with mean $1/\lambda_i$, then we can take $\mu_i=1/\lambda_i$, and $h_i$ will be exponentially distributed with mean $1$, satisfying Assumption \ref{fading_assumption}.

Consider first the multi-sensor diversity scheme. We rewrite $\max(g_1, \dots, g_M)=\max(\mu_1 h_1, \dots, \mu_M h_M)$, and then we have
$$ \max(\mu_{min} h_1, \dots, \mu_{min} h_M) \leq \max(\mu_1 h_1, \dots, \mu_M h_M) \leq \max(\mu_{max} h_1, \dots, \mu_{max} h_M)$$

We may then obtain for Rayleigh fading the bound 
\begin{equation*}
\begin{split}
\frac{\sigma_\theta^2 \sigma_{v}^2}{\sigma_\theta^2+\sigma_{v}^2} & \left[1+ \frac{\sigma_n^2 \sigma_\theta^2}{\sigma_{v}^2(\sigma_\theta^2+\sigma_{v}^2)}  \frac{1}{\mu_{max} \ln(M) }\right] (1+o(1)) \\ & \leq \mathbb{E}[D] \leq \frac{\sigma_\theta^2 \sigma_{v}^2}{\sigma_\theta^2+\sigma_{v}^2} \left[1+ \frac{\sigma_n^2 \sigma_\theta^2}{\sigma_{v}^2(\sigma_\theta^2+\sigma_{v}^2)}  \frac{1}{\mu_{min} \ln(M) }\right] (1+o(1)) 
\end{split}
\end{equation*}
The upper and lower bounds both converge to $\sigma_\theta^2 \sigma_v^2/(\sigma_\theta^2+\sigma_v^2)$ at the rate $1/\ln{M}$. So for non-i.i.d. fading channels satisfying Assumption \ref{fading_assumption} and identical sensor noise variances, the scaling behaviour of Section \ref{asymptotic_sec} is preserved in the multi-sensor diversity scheme. 

For the channel-aware ALOHA scheme, we may similarly obtain for Rayleigh fading the bound 
\begin{equation*}
\begin{split}
\sigma_\theta^2(1-\frac{1}{e}) + \frac{1}{e} \frac{\sigma_\theta^2 \sigma_{v}^2}{\sigma_\theta^2+\sigma_{v}^2} & \left[1+ \frac{\sigma_n^2 \sigma_\theta^2}{\sigma_{v}^2(\sigma_\theta^2+\sigma_{v}^2)}  \frac{1}{\mu_{max} \ln(M) }\right] (1+o(1)) \\ & \leq \mathbb{E}[D] \leq \sigma_\theta^2(1-\frac{1}{e}) + \frac{1}{e} \frac{\sigma_\theta^2 \sigma_{v}^2}{\sigma_\theta^2+\sigma_{v}^2} \left[1+ \frac{\sigma_n^2 \sigma_\theta^2}{\sigma_{v}^2(\sigma_\theta^2+\sigma_{v}^2)}  \frac{1}{\mu_{min} \ln(M) }\right] (1+o(1)) 
\end{split}
\end{equation*}
which also preserves the scaling behaviour of Section \ref{asymptotic_sec}. 

For the multi-access scheme, we will have
$$\frac{\left(\sum_{i=1}^M \sqrt{\mu_{min} h_i} \alpha_i  \right)^2}{\sum_{i=1}^M \mu_{max} h_i \alpha_i^2 \sigma_v^2 + \sigma_n^2} \leq \frac{\left(\sum_{i=1}^M \sqrt{\mu_i h_i} \alpha_i  \right)^2}{\sum_{i=1}^M \mu_i h_i \alpha_i^2 \sigma_v^2 + \sigma_n^2} \leq \frac{\left(\sum_{i=1}^M \sqrt{\mu_{max} h_i} \alpha_i  \right)^2}{\sum_{i=1}^M \mu_{min} h_i \alpha_i^2 \sigma_v^2 + \sigma_n^2}$$
and by similar calculations to Section \ref{asymptotic_sec} we may obtain the bound
$$\frac{\sigma_v^2 \mu_{min} \mathbb{E} [h_1] + \sigma_n^2}{M \mu_{max} (\mathbb{E}[\sqrt{h_1}])^2 } (1+o(1)) \leq \mathbb{E}[D] \leq \frac{\sigma_v^2 \mu_{max} \mathbb{E} [h_1] + \sigma_n^2}{M \mu_{min} (\mathbb{E} [\sqrt{h_1}])^2 } (1+o(1))$$
The upper and lower bounds both converge to 0 at the rate $1/M$, preserving the scaling behaviour of Section \ref{asymptotic_sec}. 

For the orthogonal scheme, we will have 
$$\sum_{i=1}^M \frac{\mu_{min} h_i \alpha_i^2}{\mu_{min} h_i \alpha_i^2 \sigma_v^2 + \sigma_n^2} \leq \sum_{i=1}^M \frac{\mu_i h_i \alpha_i^2}{\mu_i h_i \alpha_i^2 \sigma_v^2 + \sigma_n^2} \leq \sum_{i=1}^M \frac{\mu_{max} h_i \alpha_i^2}{\mu_{max} h_i \alpha_i^2 \sigma_v^2 + \sigma_n^2}$$
by making use of the fact that 
$$ \frac{\mu_i h_i \alpha_i^2}{\mu_i h_i \alpha_i^2 \sigma_v^2 + \sigma_n^2}$$
is an increasing function of $\mu_i$. 
We may then obtain the bounds for Rayleigh fading:
$$\frac{1}{\frac{1}{\sigma_\theta^2} +\frac{\mu_{max}}{ \sigma_n^2}} + \frac{2 \sigma_v^2 \mu_{max}^2}{M  \sigma_n^4 (\frac{1}{\sigma_\theta^2} +\frac{\mu_{max}}{\sigma_n^2})^2} (1+o(1)) \leq \mathbb{E}[D] \leq \frac{1}{\frac{1}{\sigma_\theta^2} +\frac{\mu_{min}}{ \sigma_n^2}} + \frac{2 \sigma_v^2 \mu_{min}^2}{M  \sigma_n^4 (\frac{1}{\sigma_\theta^2} +\frac{\mu_{min}}{\sigma_n^2})^2} (1+o(1)) $$
However, since the upper and lower bounds have different limits as $M \rightarrow \infty$, little more can be said in general. 

\subsection{General sensor noise variances and non-identically distributed fading channels}
By combining the results in the previous two subsections, it is clear that if we allow for both general sensor noise variances and non-identically distributed fading channels satisfying Assumption \ref{fading_assumption}, then only the multi-access scheme will preserve the scaling behaviour of Section \ref{asymptotic_sec}. 

\section{Optimal power allocation}
\label{optim_sec}
In this section we consider optimal power allocation for the multi-sensor diversity and channel-aware ALOHA schemes. For notational simplicity, and since we are also interested in the performance using optimal power allocation for large numbers of sensors, we will consider symmetric sensor networks, although the results can be generalized to general parameters such as unequal sensor noise variances and/or non-identical fading distributions as considered in the previous section.  Numerical results will show that the difference in performance between optimal power allocation and the constant power allocation used in Section \ref{asymptotic_sec} is very small. Indeed, we will argue that asymptotically the results are equivalent. Optimal power allocation for multi-access and orthogonal access schemes, with slightly different objectives and constraints, has previously been studied in \cite{Xiao_coherent_TSP} and \cite{Cui_TSP} respectively and will not be considered here. 

\subsection{Multi-sensor diversity scheme}
We are interested in minimizing the expected distortion $\mathbb{E}[D]$ subject  to an average power constraint $\mathcal{P}$. 
For the multi-sensor diversity scheme we can write this as 
\begin{equation}
\label{diversity_optim_prob}
\begin{split}
& \min_{\alpha_{i^*}^2} \mathbb{E}[D] = \min_{\alpha_{i^*}^2}  \frac{\sigma_\theta^2 \sigma_v^2}{\sigma_\theta^2 + \sigma_v^2} \mathbb{E} \left[ 1 + \frac{\frac{\sigma_n^2 \sigma_\theta^2}{\sigma_v^2(\sigma_\theta^2+\sigma_v^2)}}{g_{max}\alpha_{i^*}^2 +\frac{\sigma_n^2}{\sigma_\theta^2+\sigma_v^2}} \right] \\
& \mbox{s.t. } \mathbb{E}[\alpha_{i^*}^2] \leq \frac{\mathcal{P}}{\sigma_\theta^2+\sigma_v^2} 
\end{split}
\end{equation}
We have the following result:
\begin{lemma}
\label{lemma_diversity}
Consider the following problem
\begin{equation*}
\begin{split}
& \min_{\alpha_{i^*}^2} \mathbb{E}\left[\frac{1}{ g_{max}\alpha_{i^*}^2 +b} \right] \\
& \mbox{s.t. } \mathbb{E}[\alpha_{i^*}^2] \leq \frac{\mathcal{P}}{\sigma_\theta^2+\sigma_v^2} 
\end{split}
\end{equation*}
The optimal solution is of the form
\begin{equation}
\label{diversity_allocation}
 \alpha_{i^*}^2 = \left\{ \begin{array}{ccc} \sqrt{\frac{1}{g_{max} \nu}} - \frac{b}{g_{max}} & , & g_{max} \geq b^2 \nu  \\ 0 & , & \textrm{otherwise} \end{array}  \right.
\end{equation}
where the Lagrange multiplier $\nu$ satisfies
\begin{equation}
\label{diversity_nu_condition}
\int_{b^2 \nu}^\infty \left( \sqrt{\frac{1}{g_{max} \nu}} - \frac{b}{g_{max}} \right) p(g_{max}) dg_{max} = \frac{\mathcal{P}}{\sigma_\theta^2+\sigma_v^2}
\end{equation}
\end{lemma}

\begin{proof}
The derivation uses similar techniques to the capacity maximization problems for fading channels in \cite{GoldsmithVaraiya},\cite{KnoppHumblet1}, and is omitted for brevity. 
\end{proof}
Using Lemma \ref{lemma_diversity}, the optimal power allocation for problem (\ref{diversity_optim_prob}) is given by (\ref{diversity_allocation}), with $b = \frac{\sigma_n^2}{\sigma_\theta^2+\sigma_v^2}$. The expected distortion under optimal power allocation can be computed as
\begin{equation}
\label{ED_diversity_optimal}
\mathbb{E}[D] = \int_{b^2 \nu}^\infty \frac{\sigma_\theta^2 \sigma_v^2}{\sigma_\theta^2+\sigma_v^2} \left( 1 + \frac{\sigma_n^2 \sigma_\theta^2}{\sigma_v^2(\sigma_\theta^2+\sigma_v^2)} \sqrt{\frac{\nu}{g_{max}}} \right) p(g_{max}) dg_{max} + \int_0^{b^2 \nu} \sigma_\theta^2 p(g_{max}) dg_{max}
\end{equation}
where $\nu$ satisfies (\ref{diversity_nu_condition}),
and can be determined numerically. 

\subsection{Channel-aware ALOHA scheme}
\label{aloha_optim_sec}
For the channel-aware ALOHA scheme, the problem of minimizing the expected distortion subject $E[D]$ to an average power constraint $\mathcal{P}$ can be written as
\begin{equation}
\label{aloha_optim_prob}
\begin{split}
& \min_{\alpha_{i}^2} \mathbb{E}[D] = \min_{\alpha_{i}^2}  \sigma_\theta^2 \left[1 - (1-\frac{1}{M})^{M-1} \right] + M (1-\frac{1}{M})^{M-1} \int_{T}^\infty \frac{\sigma_\theta^2 \sigma_v^2}{\sigma_\theta^2+\sigma_v^2} \left( 1 + \frac{\frac{\sigma_n^2 \sigma_\theta^2}{\sigma_v^2(\sigma_\theta^2+\sigma_v^2)}}{\alpha_i^2 g_i+\frac{\sigma_n^2}{\sigma_\theta^2+\sigma_v^2}} \right) p(g_i) dg_i \\
& \mbox{s.t. }  M \int_T^\infty \alpha_i^2 p(g_i) dg_i  \leq \frac{\mathcal{P}}{\sigma_\theta^2+\sigma_v^2} 
\end{split}
\end{equation}
Similar to the multi-sensor diversity scheme, we have the following result:
\begin{lemma}
\label{lemma_distributed}
Consider the following problem
\begin{equation*}
\begin{split}
& \min_{\alpha_{i}^2} \int_T^\infty \frac{1}{ g_{i}\alpha_{i}^2 +b} p(g_i)dg_i \\
& \mbox{s.t. } \int_T^\infty \alpha_i^2 p(g_i) dg_i \leq \frac{\mathcal{P}}{M(\sigma_\theta^2+\sigma_v^2)} 
\end{split}
\end{equation*}
The optimal solution is of the form
\begin{equation}
\label{aloha_allocation}
 \alpha_{i}^2 = \left\{ \begin{array}{ccc} \sqrt{\frac{1}{g_i \nu}} - \frac{b}{g_i} & , & g_i \geq \max(T,b^2 \nu)  \\ 0 & , & \textrm{otherwise} \end{array}  \right.
\end{equation}
where the Lagrange multiplier $\nu$ satisfies
\begin{equation}
\label{aloha_nu_condition}
\!\!\!\!\!\!\!\!\!\! \int\limits_{\max(T,b^2 \nu)}^\infty \!\!\!\!\!\!\!\! \left( \sqrt{\frac{1}{g_{i} \nu}} - \frac{b}{g_{i}} \right) p(g_{i}) dg_{i} = \frac{\mathcal{P}}{M(\sigma_\theta^2+\sigma_v^2)}
\end{equation}
\end{lemma}

Using Lemma \ref{lemma_distributed}, the optimal power allocation for problem (\ref{aloha_optim_prob}) is given by (\ref{aloha_allocation}), with $b = \frac{\sigma_n^2}{\sigma_\theta^2+\sigma_v^2}$. The expected distortion under optimal power allocation can be computed as
\begin{equation}
\label{ED_aloha_optimal}
\begin{split}
\mathbb{E}[D]   = & \sigma_\theta^2 \left[1 - (1-\frac{1}{M})^{M-1} \right] + M(1-\frac{1}{M})^{M-1} \!\!\!\!\!\!\!\!\!\! \int\limits_{\max(T,b^2 \nu)}^\infty \!\!\!\!\!\!\!\! \frac{\sigma_\theta^2 \sigma_v^2}{\sigma_\theta^2+\sigma_v^2} \left( 1 + \frac{\sigma_n^2 \sigma_\theta^2}{\sigma_v^2(\sigma_\theta^2+\sigma_v^2)}\sqrt{\frac{\nu}{g_i}} \right) p(g_i) dg_i \\ & + M(1 - \frac{1}{M})^{M-1}  \!\!\!\!\!\!\!\!\!\! \int\limits_T^{\max(T,b^2 \nu)} \!\!\!\!\!\!\!\! \sigma_\theta^2 p(g_i) dg_i
\end{split}
\end{equation}
where $\nu$ satisfies (\ref{aloha_nu_condition}).

\subsection{Numerical studies}
\label{optimal_numerical_sec}
We again consider a situation with $\sigma_\theta^2=1, \sigma_v^2=0.2, \sigma_n^2=0.1$, and let $g_i, \forall i$ be exponentially distributed with mean $1/2$. For a fair comparison with the results of Section \ref{numerical_sec}, when performing optimal power allocation we will take $\mathcal{P}/(\sigma_\theta^2 + \sigma_v^2)=1$. 

\begin{figure}[tbp]
\centering
\includegraphics[width=8.0cm]{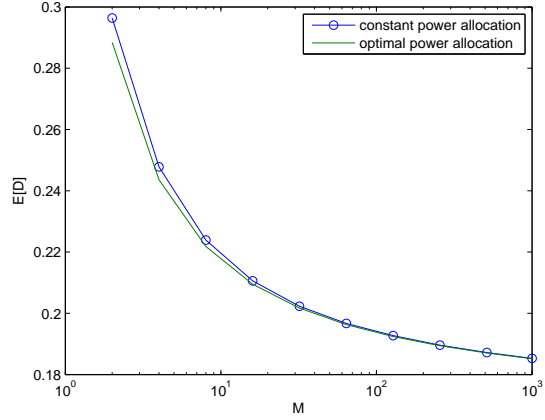}
\caption{Multi-sensor diversity scheme. Comparison between constant and optimal power allocation}
\label{diversity_optimal_plot}
\end{figure}

\begin{figure}[tbp]
\centering
\includegraphics[width=8.0cm]{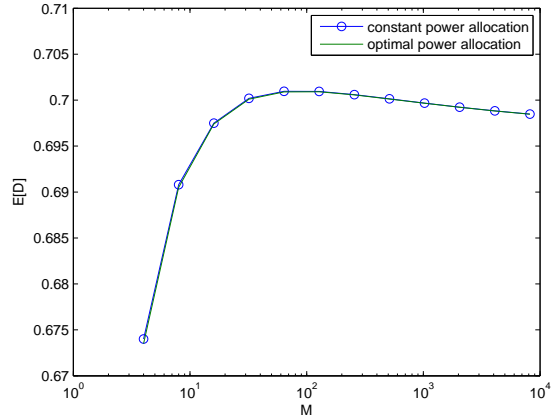}
\caption{Channel-aware ALOHA scheme. Comparison between constant and optimal power allocation}
\label{aloha_optimal_plot}
\end{figure}

In Fig. \ref{diversity_optimal_plot} we plot the expected distortion under constant (when the sensor is transmitting) and optimal power allocation, for the multi-sensor diversity scheme with different numbers of sensors. 
In Fig. \ref{aloha_optimal_plot} we plot the expected distortion under constant and optimal power allocation, for the channel-aware ALOHA scheme with different numbers of sensors. 
The performance using constant power allocation can be seen to be very close to the performance under optimal power allocation, particularly for large numbers of sensors. In the next subsection we will attempt to explain this phenomenon. 

\subsection{Asymptotic behaviour under optimal power allocation}
In this subsection we will prove why the optimal power allocation and constant power allocation schemes perform so close to each other, 
especially for large $M$. We will assume that $g_i$ are exponentially distributed with mean $1/\lambda$. We will also take $\mathcal{P}/(\sigma_\theta^2 + \sigma_v^2)=1$. 

\subsubsection{Multi-sensor diversity scheme}
Before we state and prove the main theorem, we first give a preliminary result.
\begin{lemma}
\label{diversity_optimal_lemma}
For the multi-sensor diversity scheme under optimal power allocation, $\nu \rightarrow 0$ as $M\rightarrow \infty$, where $\nu$ satisfies (\ref{diversity_nu_condition}). 
\end{lemma}
See Appendix \ref{diversity_optimal_lemma_appendix} for the proof of Lemma \ref{diversity_optimal_lemma}. 
We will now prove the following:
\begin{theorem}
For the multi-sensor diversity scheme under optimal power allocation,
$$\mathbb{E}[D] \sim \frac{\sigma_\theta^2 \sigma_v^2}{\sigma_\theta^2+\sigma_v^2} \left[1+ \frac{\sigma_n^2 \sigma_\theta^2}{\sigma_v^2(\sigma_\theta^2+\sigma_v^2)}  \frac{\lambda}{\ln(M) }\right]
 \textrm{ as } M \rightarrow \infty.$$
\end{theorem}

\begin{proof}
Firstly, by using similar techniques to Appendix \ref{gmax_lemma_appendix}, we can derive that
\begin{equation}
\label{asympt_integral_1}
\int_{0}^\infty \frac{1}{\sqrt{x}} M (1-e^{-\lambda x})^{M-1} \lambda e^{-\lambda x} dx \sim  \sqrt{\frac{ \lambda}{\ln(M)}} 
\end{equation}
and 
\begin{equation}
\label{asympt_integral_2}
\int_{0}^\infty \frac{1}{x} M (1-e^{-\lambda x})^{M-1} \lambda e^{-\lambda x} dx  \sim \frac{\lambda}{\ln(M)}
\end{equation}
By Lemma \ref{diversity_optimal_lemma} and (\ref{asympt_integral_1})-(\ref{asympt_integral_2}), the condition 
$$ \int_{b^2 \nu}^\infty \left( \sqrt{\frac{1}{g_{max} \nu}} - \frac{b}{g_{max}} \right) p(g_{max}) dg_{max} = 1$$
is asymptotically
$$\sqrt{\frac{\lambda}{\nu \ln(M)}} - \frac{b \lambda}{\ln(M)} \sim 1$$
We can easily solve for $\nu$ to get
\begin{equation}
\label{diversity_nu_asymptotic}
 \nu \sim \frac{\lambda}{\ln(M)} \frac{1}{1 + \frac{2b\lambda}{\ln(M)}+(\frac{b\lambda}{\ln(M)})^2} \sim \frac{\lambda}{\ln(M)}
\end{equation}
and so
$$\int_{b^2 \nu}^\infty \sqrt{\frac{\nu}{x}} M (1-e^{-\lambda x})^{M-1} \lambda e^{-\lambda x} dx \sim \sqrt{\frac{\nu \lambda}{\ln(M)}} \sim \frac{\lambda}{\ln(M)}$$ 
We also note that
\begin{equation*}
\begin{split}
(1-e^{-\lambda b^2 \nu})^M & = (1-\exp(-\frac{\lambda^2 b^2}{\ln(M)}))^M \\
& = O \left(\left(\frac{\lambda^2 b^2}{\ln(M)} \right)^M \right) \\
& = o\left(\frac{1}{\ln(M)} \right)
\end{split}
\end{equation*}
Hence from (\ref{ED_diversity_optimal}),
\begin{equation*}
\begin{split}
\mathbb{E}[D] & \sim \frac{\sigma_\theta^2 \sigma_v^2}{\sigma_\theta^2+\sigma_v^2} \left[1+ \frac{\sigma_n^2 \sigma_\theta^2}{\sigma_v^2(\sigma_\theta^2+\sigma_v^2)}  \frac{\lambda}{\ln(M) }\right] + \sigma_\theta^2 (1-e^{-\lambda b^2 \nu})^M \\
& \sim \frac{\sigma_\theta^2 \sigma_v^2}{\sigma_\theta^2+\sigma_v^2} \left[1+ \frac{\sigma_n^2 \sigma_\theta^2}{\sigma_v^2(\sigma_\theta^2+\sigma_v^2)}  \frac{\lambda}{\ln(M) }\right]
\end{split}
\end{equation*}
which is the same asymptotic expression as (\ref{diversity_asympt}) of Section \ref{asymptotic_sec}. 
\end{proof}

\subsubsection{Channel-aware ALOHA scheme}
We again first give a preliminary result before stating and proving the main theorem.
\begin{lemma}
\label{aloha_optimal_lemma}
For the channel-aware ALOHA scheme under optimal power allocation, 
 $T > b^2 \nu$ for $M$ sufficiently large, where $T=\frac{1}{\lambda} \ln(M)$ and $\nu$ satisfies (\ref{aloha_nu_condition}). 
\end{lemma}
See Appendix \ref{aloha_optimal_lemma_appendix} for the proof of Lemma \ref{aloha_optimal_lemma}. 
We will now prove the following:
\begin{theorem}
For the channel-aware ALOHA scheme under optimal power allocation, 
$$\mathbb{E}[D] \sim \sigma_\theta^2 (1-\frac{1}{e}) + \frac{1}{e} \frac{\sigma_\theta^2 \sigma_v^2}{\sigma_\theta^2+\sigma_v^2} \left[1 + \frac{\sigma_n^2 \sigma_\theta^2}{\sigma_v^2(\sigma_\theta^2+\sigma_v^2)} \frac{\lambda}{\ln M } \right]
\textrm{ as }M \rightarrow \infty.$$
\end{theorem}
 
\begin{proof}
Recall that $T=\frac{1}{\lambda} \ln(M)$. 
First note that we can compute the following integrals:
\begin{equation}
\label{aloha_integral_1}
\int_T^\infty \frac{1}{\sqrt{x}} \lambda e^{-\lambda x} dx = \sqrt{ \lambda \pi} \textrm{erfc}(\sqrt{\lambda T}) = \sqrt{\lambda\pi} \textrm{erfc}(\sqrt{\ln(M) })
\end{equation}
and
\begin{equation}
\label{aloha_integral_2}
\int_T^\infty \frac{1}{x} \lambda e^{-\lambda x} dx = \lambda E_1 (\lambda T) = \lambda E_1 (\ln(M))
\end{equation}
By Lemma \ref{aloha_optimal_lemma} and (\ref{aloha_integral_1})-(\ref{aloha_integral_2}), the condition
$$ \!\!\!\!\!\!\!\!\!\! \int\limits_{\max(T,b^2 \nu)}^\infty \!\!\!\!\!\!\!\!  \left( \sqrt{\frac{1}{g_{i} \nu}} - \frac{b}{g_{i}} \right) p(g_{i}) dg_{i} = \frac{1}{M}$$ 
is asymptotically 
$$ \sqrt{\frac{\lambda \pi}{\nu}} \textrm{erfc}(\sqrt{\ln(M)}) - b \lambda E_1(\ln(M)) \sim \frac{1}{M}$$
and so
\begin{equation*}
\begin{split}
 \sqrt{\nu} & \sim \frac{\sqrt{\lambda\pi} \textrm{erfc}(\sqrt{\ln(M)})}{1/M+b\lambda E_1(\ln(M))} \\
 & \sim \frac{\sqrt{\lambda\pi}  \frac{e^{-\ln(M)}}{\sqrt{\pi \ln(M)}}}{1/M + b \lambda \frac{e^{-\ln(M)}}{\ln(M)}} \\
 & \sim \sqrt{\frac{\lambda}{\ln(M)}}
\end{split}
\end{equation*}
Then 
\begin{equation*}
\begin{split}
 \int\limits_{\max(T,b^2 \nu)}^\infty \!\!\!\!\!\!\!\! \sqrt{\frac{\nu}{x}} \lambda e^{-\lambda x} dx & =   \int_T^\infty \sqrt{\frac{\nu}{x}} \lambda e^{-\lambda x} dx \\
 & = \sqrt{\nu\lambda\pi} \textrm{erfc}(\sqrt{\ln(M) }) \\
 & \sim \lambda \sqrt{\frac{\pi}{\ln(M)}} \frac{e^{-\ln(M)}}{\sqrt{\pi \ln(M)}}\\
 & = \frac{\lambda}{M \ln(M)}
\end{split}
\end{equation*}
Hence from (\ref{ED_aloha_optimal}) we have 
\begin{equation*}
\begin{split}
\mathbb{E}[D] & \sim \sigma_\theta^2 (1-\frac{1}{e}) + M \frac{1}{e} \frac{\sigma_\theta^2 \sigma_v^2}{\sigma_\theta^2+\sigma_v^2} \left[\frac{1}{M} + \frac{\sigma_n^2 \sigma_\theta^2}{\sigma_v^2(\sigma_\theta^2+\sigma_v^2)} \frac{\lambda}{M \ln M } \right] \\
& = \sigma_\theta^2 (1-\frac{1}{e}) + \frac{1}{e} \frac{\sigma_\theta^2 \sigma_v^2}{\sigma_\theta^2+\sigma_v^2} \left[1 + \frac{\sigma_n^2 \sigma_\theta^2}{\sigma_v^2(\sigma_\theta^2+\sigma_v^2)} \frac{\lambda}{\ln M } \right]
\end{split}
\end{equation*}
which is the same asymptotic expression as (\ref{aloha_asympt}) of Section \ref{asymptotic_sec}. 
\end{proof}

\section{Optimal threshold selection for channel-aware ALOHA scheme}
\label{threshold_sec}
So far in this paper we have used the choice of threshold $T=\frac{1}{\lambda} \ln(M)$ in the channel-aware ALOHA scheme. 
In this section we will consider the optimal choice of threshold in the channel-aware ALOHA scheme for symmetric sensor networks. We consider both threshold optimization under constant power allocation, and a joint threshold/power optimization. We will assume Rayleigh fading, so that $g_i$ are exponentially distributed with mean $1/\lambda$. 

\subsection{Optimal thresholds under constant power allocation}
Recall that in the channel-aware ALOHA scheme each sensor transmits when $g_i > T$. The problem we now consider is to determine the optimal choice of $T$ to minimize the expected distortion. 

Note that under Rayleigh fading, $\Pr(g_i > T) = e^{-\lambda T}$. For a fair comparison with the model of Section \ref{aloha_model_sec} we will normalise the powers, and let $\alpha_i$ be of the form 
$$ \alpha_i^2 = \frac{e^{\lambda T}}{M} $$
The expected distortion can then be derived similar to Section \ref{aloha_asymptotic_sec} as 
\begin{equation}
\label{ED_optim_threshold}
\begin{split}
\mathbb{E}[D]  & =  \sigma_\theta^2 \left[ 1-M e^{-\lambda T} (1-e^{-\lambda T})^{M-1} \right] + M(1-e^{-\lambda T})^{M-1} \int_T^\infty \left(\frac{1}{\sigma_\theta^2} + \frac{g_i e^{\lambda T}/M}{g_i \sigma_v^2 e^{\lambda T}/M+\sigma_n^2} \right)^{-1} p(g_i) dg_i \\
& =  \sigma_\theta^2 \left[ 1-M e^{-\lambda T} (1-e^{-\lambda T})^{M-1} \right] \\ &  + M(1-e^{-\lambda T})^{M-1} \frac{\sigma_\theta^2 \sigma_v^2}{\sigma_\theta^2+\sigma_v^2} \left[ e^{-\lambda T} + \frac{\sigma_n^2 \sigma_\theta^2 M e^{-\lambda T}}{\sigma_v^2 (\sigma_\theta^2+\sigma_v^2)} \lambda \exp\left( \frac{\lambda \sigma_n^2 M e^{-\lambda T}}{\sigma_\theta^2 + \sigma_v^2}\right) E_1 \left(\lambda \left(\frac{\sigma_n^2 M e^{-\lambda T}}{\sigma_\theta^2 + \sigma_v^2} + T\right) \right)\right]
\end{split}
\end{equation}
The optimal threshold can then be found by numerically searching for the $T^*$ that satisfies $\frac{d \mathbb{E}[D]}{dT}|_{T=T^*} = 0$ and $\frac{d^2\mathbb{E}[D]}{dT^2}|_{T=T^*} > 0$.

\subsection{Joint threshold/power optimization}
Here we wish to optimize both the threshold and determine the optimal power allocation that will minimize the expected distortion, subject to an average power constraint. The problem can be written as 
\begin{equation}
\label{aloha_joint_optim_prob}
\begin{split}
& \min_{T, \alpha_{i}^2}  \sigma_\theta^2 \left[ 1-M e^{-\lambda T} (1-e^{-\lambda T})^{M-1} \right] + M(1-e^{-\lambda T})^{M-1} \int_T^\infty \frac{\sigma_\theta^2 \sigma_v^2}{\sigma_\theta^2+\sigma_v^2} \left( 1 + \frac{\frac{\sigma_n^2 \sigma_\theta^2}{\sigma_v^2(\sigma_\theta^2+\sigma_v^2)}}{\alpha_i^2 g_i+\frac{\sigma_n^2}{\sigma_\theta^2+\sigma_v^2}} \right) p(g_i) dg_i \\
& \mbox{s.t. }  M \int_T^\infty \alpha_i^2 p(g_i) dg_i  \leq \frac{\mathcal{P}}{\sigma_\theta^2+\sigma_v^2} 
\end{split}
\end{equation}
To solve (\ref{aloha_joint_optim_prob}), we note that for a given $T$, it can be shown similar to Section \ref{aloha_optim_sec} that the optimal power allocation has the form 
\begin{equation*}
%\label{aloha_allocation}
 \alpha_{i}^2 = \left\{ \begin{array}{ccc} \sqrt{\frac{1}{g_i \nu}} - \frac{b}{g_i} & , & g_i \geq \max(T,b^2 \nu)  \\ 0 & , & \textrm{otherwise} \end{array}  \right.
\end{equation*}
where $\nu$ satisfies
\begin{equation*}
%\label{aloha_nu_condition}
\!\!\!\!\!\!\!\!\!\! \int\limits_{\max(T,b^2 \nu)}^\infty \!\!\!\!\!\!\!\! \left( \sqrt{\frac{1}{g_{i} \nu}} - \frac{b}{g_{i}} \right) p(g_{i}) dg_{i} = \frac{\mathcal{P}}{M(\sigma_\theta^2+\sigma_v^2)}
\end{equation*}
and 
\begin{equation}
\label{ED_optim_joint}
\begin{split}
\mathbb{E}[D]   = & \sigma_\theta^2 \left[1 - M e^{-\lambda T}(1-e^{-\lambda T})^{M-1} \right] + M(1-e^{-\lambda T})^{M-1} \!\!\!\!\!\!\!\!\!\! \int\limits_{\max(T,b^2 \nu)}^\infty \!\!\!\!\!\!\!\! \frac{\sigma_\theta^2 \sigma_v^2}{\sigma_\theta^2+\sigma_v^2} \left( 1 + \frac{\sigma_n^2 \sigma_\theta^2}{\sigma_v^2(\sigma_\theta^2+\sigma_v^2)}\sqrt{\frac{\nu}{g_i}} \right) p(g_i) dg_i \\ & + M(1 - e^{-\lambda T})^{M-1}  \!\!\!\!\!\!\!\!\!\! \int\limits_T^{\max(T,b^2 \nu)} \!\!\!\!\!\!\!\! \sigma_\theta^2 p(g_i) dg_i
\end{split}
\end{equation}
With this, we may then again perform a line search to find the optimal $T^*$ that minimizes $\mathbb{E}[D]$. 

\subsection{Numerical studies}
We again consider the situation with $\sigma_\theta^2=1, \sigma_v^2=0.2, \sigma_n^2=0.1$, and let $g_i, \forall i$ be exponentially distributed with mean $1/2$. 

\begin{figure}[tbp]
\centering
\includegraphics[width=8.0cm]{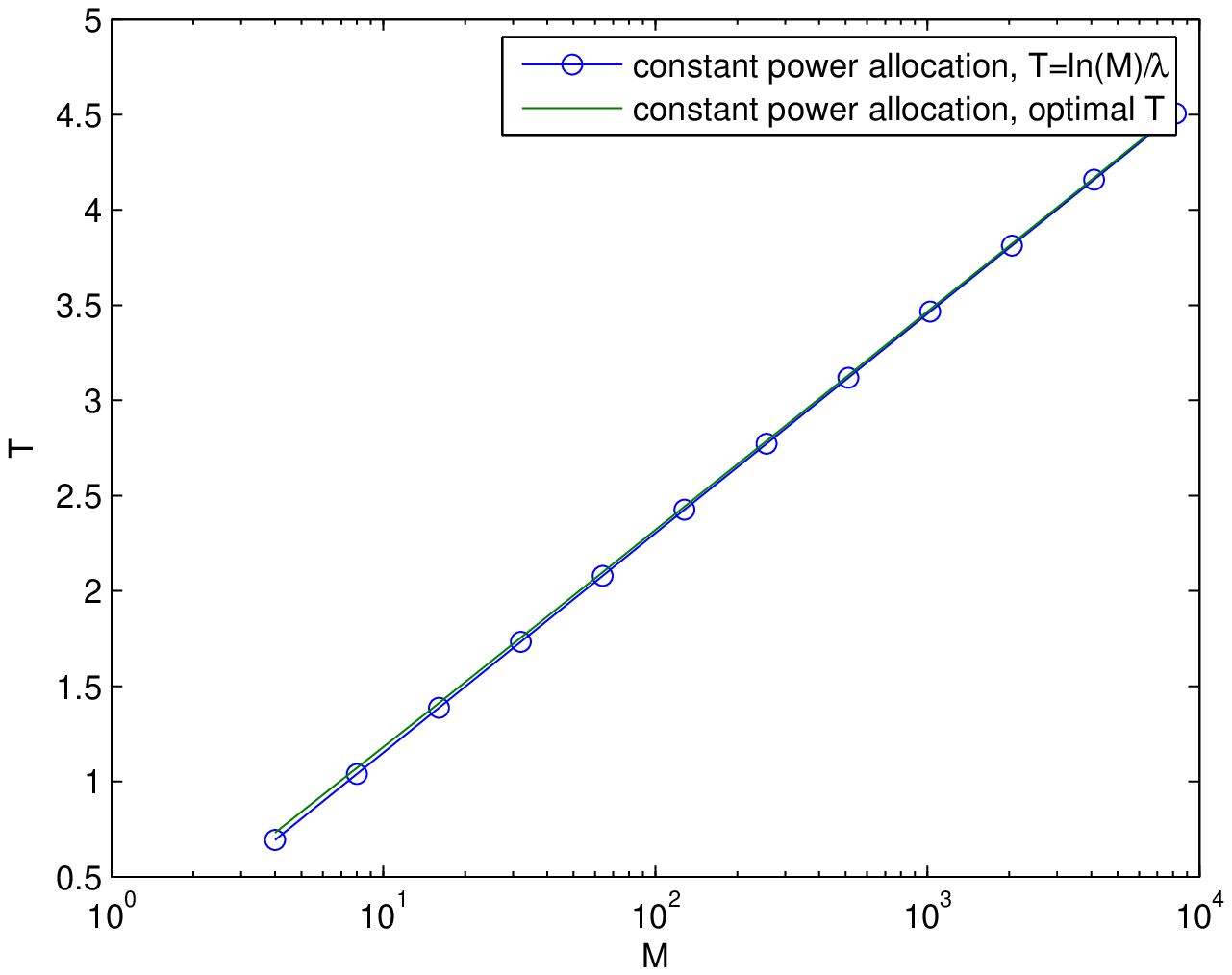}
\caption{Constant power allocation. Simple and optimal thresholding}
\label{aloha_threshold_plot2}
\end{figure}

\begin{figure}[tbp]
\centering
\includegraphics[width=8.0cm]{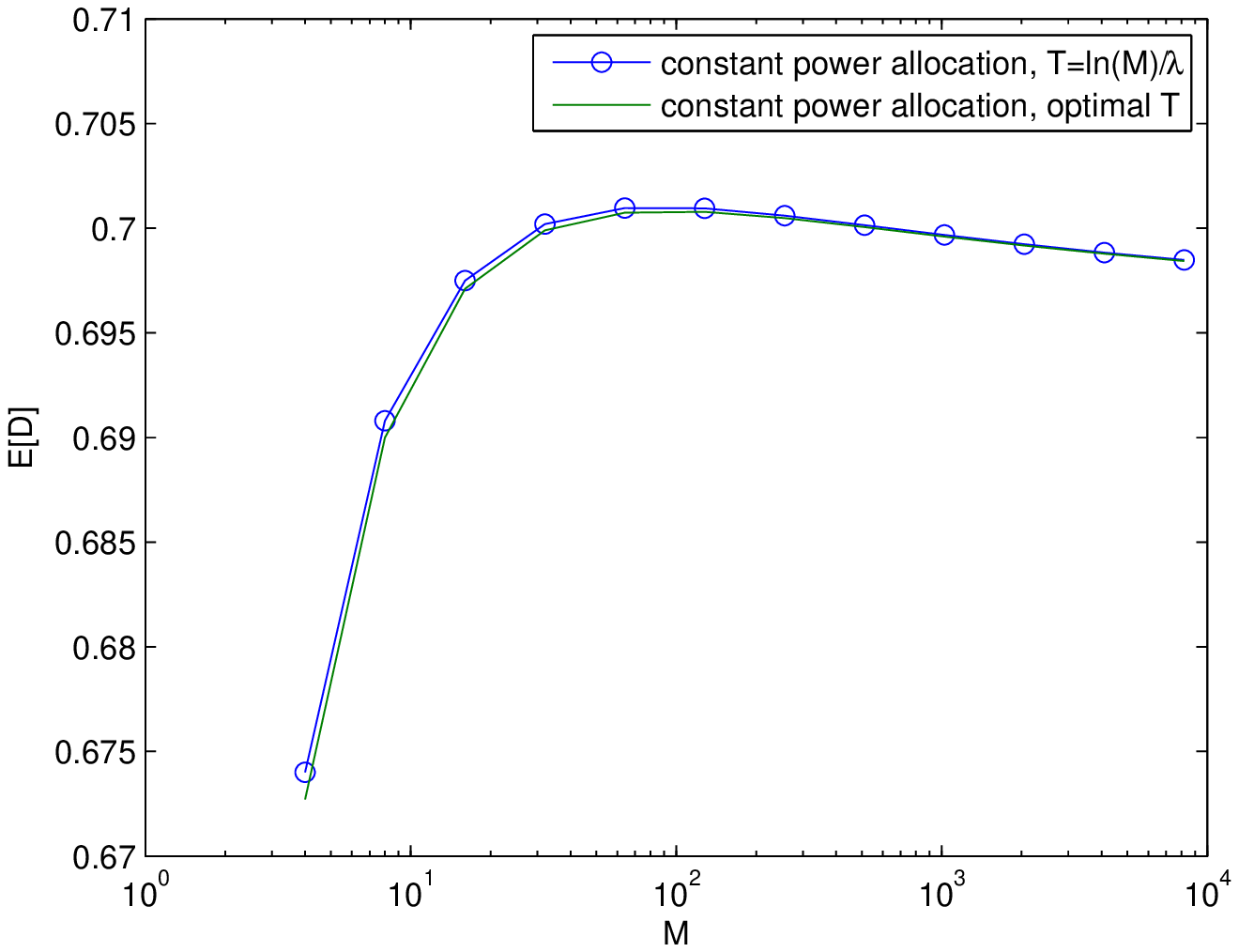}
\caption{Constant power allocation. Simple and optimal thresholding}
\label{aloha_threshold_plot}
\end{figure}

In Figures \ref{aloha_threshold_plot2} and \ref{aloha_threshold_plot} we plot the thresholds and expected distortion under constant power allocation, comparing the performance using optimal thresholds and the simple choice of threshold $T=\frac{1}{\lambda} \ln(M)$. The results can be seen to be very close to each other. 

\begin{figure}[tbp]
\centering
\includegraphics[width=8.0cm]{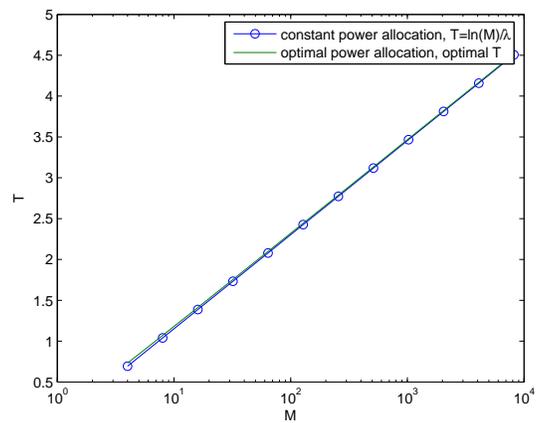}
\caption{Constant power allocation and simple thresholding vs optimal power allocation and optimal thresholding}
\label{aloha_optimal_joint_plot2}
\end{figure}

\begin{figure}[tbp]
\centering
\includegraphics[width=8.0cm]{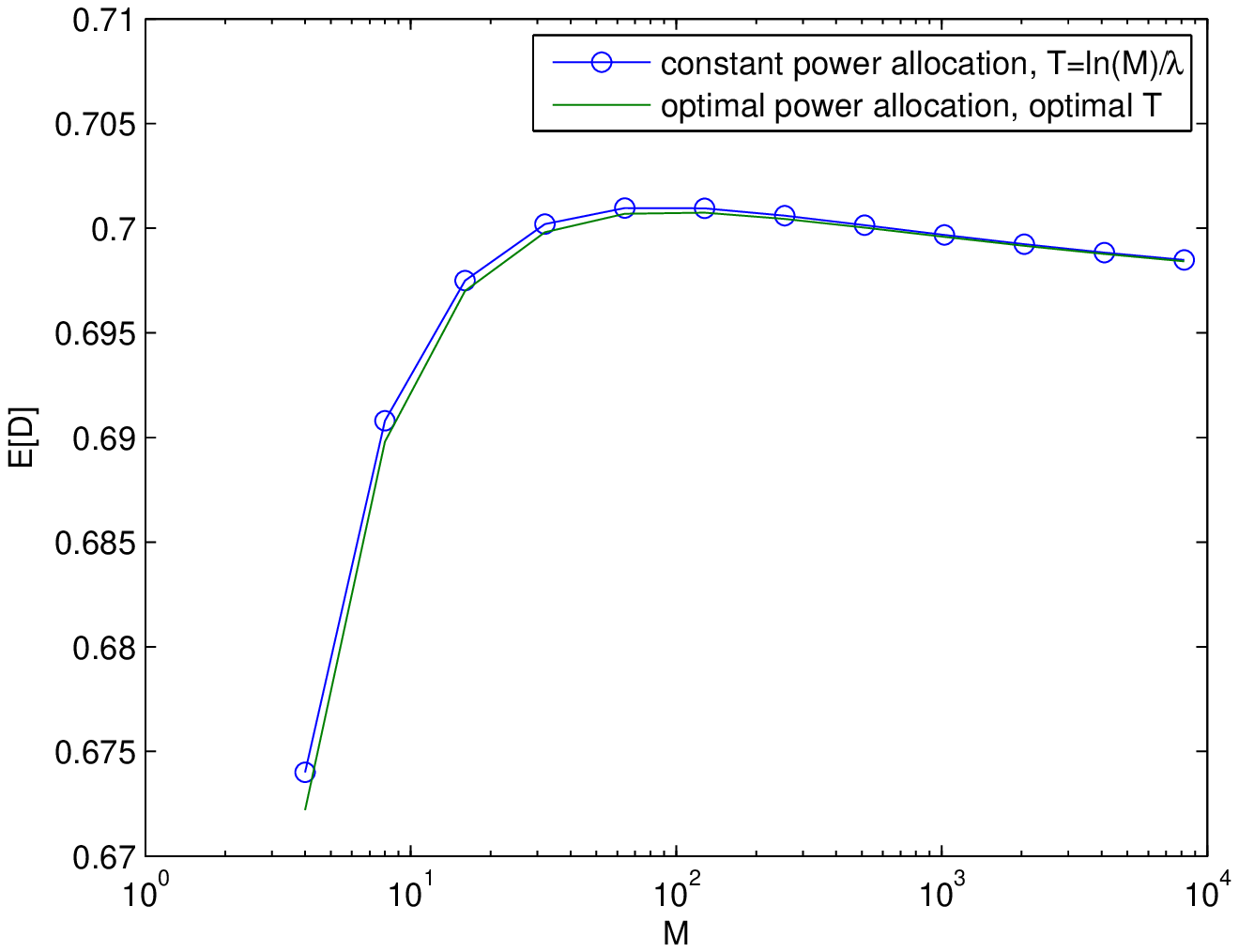}
\caption{Constant power allocation and simple thresholding vs optimal power allocation and optimal thresholding}
\label{aloha_optimal_joint_plot}
\end{figure}

In Figures \ref{aloha_optimal_joint_plot2} and \ref{aloha_optimal_joint_plot} we plot the thresholds and expected distortion, comparing the performance using optimal power allocation with optimal thresholds, and constant power allocation with the simple threshold $T=\frac{1}{\lambda} \ln(M)$. The results can also be seen to be very close to each other. 

\subsection{Optimal thresholding for large $M$}
From the numerical results in the previous subsection, it appears that the optimal thresholds (under both constant power and optimal power allocation) are asymptotically equal to $\frac{1}{\lambda} \ln(M)$. Indeed, we have the following result:

\begin{lemma}
Under both constant power and optimal power allocation, the optimal thresholds $T^*$ satisfy $T^* \sim \frac{1}{\lambda} \ln(M)$ as $M \rightarrow \infty$. 
\end{lemma}

\begin{proof}
Regard $T$ as a function of $M$. Consider the term 
$$M e^{-\lambda T(M)} (1-e^{-\lambda T(M)})^{M-1}$$
in both the expressions (\ref{ED_optim_threshold}) and (\ref{ED_optim_joint}). 
If the thresholds are chosen such that this term decays to zero as $M \rightarrow \infty$, then in both (\ref{ED_optim_threshold}) and (\ref{ED_optim_joint}) we have $\sigma_\theta^2[1-M e^{-\lambda T(M)} (1-e^{-\lambda T(M)})^{M-1}] \rightarrow \sigma_\theta^2$, and hence 
 $$\mathbb{E}[D] \geq \sigma_\theta^2 (1+o(1)) \textrm{ as } M \rightarrow \infty.$$ However, we already know from Section \ref{aloha_asymptotic_sec} that the choice $T(M) = \frac{1}{\lambda} \ln(M)$ results in a lower expected distortion than this. Thus a necessary condition for the optimal choice of thresholds $T^* (M)$ is that the term $M e^{-\lambda T^*(M)} (1-e^{-\lambda T^*(M)})^{M-1}$ does not converge to zero as $M \rightarrow \infty$. 

Now for the term $M e^{-\lambda T^*(M)}$ to not converge to zero, one needs $T^*(M) \leq \frac{1}{\lambda} \ln(M) (1+o(1))$. For the term $(1-e^{-\lambda T^*(M)})^{M-1}$ to not converge to zero, one needs $T^*(M) \geq \frac{1}{\lambda} \ln(M) (1+o(1))$. Combining these two statements, one then gets that the optimal thresholds must have the form $T^*(M) \sim \frac{1}{\lambda} \ln(M)$. 
\end{proof}

Intuitively, one could next attempt to substitute $T^* \sim \frac{1}{\lambda} \ln(M)$ into (\ref{ED_optim_threshold}) or (\ref{aloha_joint_optim_prob}) in order to obtain the asymptotic expression (\ref{aloha_asympt}) for the expected distortion. This however is not a rigorous argument since performing the operation $e^{-\lambda T^*}$ does not retain the asymptotic relation $\sim$. 
We can prove however, the following  weaker result:
\begin{lemma}
\label{optimal_threshold_lemma}
Under both constant power and optimal power allocation, and optimal thresholding, we have
$$\left( \sigma_\theta^2 (1-\frac{1}{e}) + \frac{1}{e} \frac{\sigma_\theta^2 \sigma_v^2}{\sigma_\theta^2+\sigma_v^2} \right)(1+o(1)) \leq \mathbb{E}[D] \leq \sigma_\theta^2 (1-\frac{1}{e}) + \frac{1}{e} \frac{\sigma_\theta^2 \sigma_v^2}{\sigma_\theta^2+\sigma_v^2} \left[1 + \frac{\sigma_n^2 \sigma_\theta^2}{\sigma_v^2(\sigma_\theta^2+\sigma_v^2)} \frac{\lambda}{\ln M } \right] (1+o(1))$$
as $M \rightarrow \infty$.
\end{lemma}

\begin{proof}
The upper bound on $\mathbb{E}[D]$ comes from the fact that the sub-optimal choice $T=\frac{1}{\lambda} \ln(M)$ with constant power allocation gives the asymptotic behaviour (\ref{aloha_asympt}) in Section \ref{aloha_asymptotic_sec}.
For the lower bound, consider the term 
$$\sigma_\theta^2 \left[ 1-M e^{-\lambda T} (1-e^{-\lambda T})^{M-1} \right]  + Me^{-\lambda T} (1-e^{-\lambda T})^{M-1} \frac{\sigma_\theta^2 \sigma_v^2}{\sigma_\theta^2+\sigma_v^2}$$
in either (\ref{ED_optim_threshold}) or (\ref{aloha_joint_optim_prob}). One can easily show that this term is minimized by using $T=\frac{1}{\lambda} \ln(M)$, resulting in
\begin{equation*}
\begin{split}
\sigma_\theta^2 &  \left[ 1-M e^{-\lambda T} (1-e^{-\lambda T})^{M-1} \right]  + Me^{-\lambda T} (1-e^{-\lambda T})^{M-1} \frac{\sigma_\theta^2 \sigma_v^2}{\sigma_\theta^2+\sigma_v^2} \\ & = \sigma_\theta^2 \left[1 - (1-\frac{1}{M})^{M-1} \right] + (1-\frac{1}{M})^{M-1} \frac{\sigma_\theta^2 \sigma_v^2}{\sigma_\theta^2+\sigma_v^2} \\ & \sim \sigma_\theta^2 (1-\frac{1}{e}) + \frac{1}{e} \frac{\sigma_\theta^2 \sigma_v^2}{\sigma_\theta^2+\sigma_v^2}
\end{split}
\end{equation*}
Hence from either (\ref{ED_optim_threshold}) or (\ref{aloha_joint_optim_prob}), $\mathbb{E}[D] \geq \left( \sigma_\theta^2 (1-\frac{1}{e}) + \frac{1}{e} \frac{\sigma_\theta^2 \sigma_v^2}{\sigma_\theta^2+\sigma_v^2} \right)(1+o(1))$.
\end{proof}
By Lemma \ref{optimal_threshold_lemma}, we see that $\mathbb{E}[D]$ will go to the same limiting value $\sigma_\theta^2 (1-\frac{1}{e}) + \frac{1}{e} \frac{\sigma_\theta^2 \sigma_v^2}{\sigma_\theta^2+\sigma_v^2}$ at a rate at least as fast as $1/\ln(M)$. However, showing that the rate is exactly $1/\ln(M)$, and that the exact asymptotic behaviour is given by (\ref{aloha_asympt}), remain open issues. 
\section{Conclusion}
The asymptotic behaviour for decentralized estimation of an i.i.d Gaussian source, using the analog amplify and forwarding technique under a number of different multiple access schemes, has been studied. Focusing on the expected distortion, the rate of decay of $1/\ln(M)$ has been shown for multi-sensor diversity and channel-aware ALOHA schemes, while the coherent multi-access and orthogonal access schemes have decay rates of $1/M$. The optimal power allocation for the multi-sensor diversity schemes has also been derived, and we have found that simple power allocation policies can actually approach the optimal results very closely as the number of sensors increases. 

The diversity schemes considered here can obviously be made more sophisticated. For instance, instead of just the best sensor transmitting their measurement to the fusion center, we could have the best $N$ sensors transmitting, with $N \geq 2$. This could be useful in particular when the sensor measurements are spatially 
correlated. For another example, in the channel-aware ALOHA scheme, instead of assuming collision when more than one sensor transmits at the same time, we might be able to combine them by coherently adding up the sensor transmissions as in the multi-access scheme. 
Analysis of these schemes will be more complicated, but could constitute possible areas of future investigation.  

\begin{appendix}
\subsection{Proof of Lemma \ref{gmax_lemma}}
\label{gmax_lemma_appendix}
\begin{proof}
The maximum of $M$ i.i.d. exponential random variables with mean $1/\lambda$, has cumulative distribution function
$$ F(x) = (1-e^{-\lambda x})^M$$
and hence the probability density function
$$ p(x) = M (1-e^{-\lambda x})^{M-1}\lambda e^{-\lambda x}.$$
We wish to find the large $M$ behaviour of
\begin{eqnarray*}
\mathbb{E}\left[ \frac{1}{X+b} \right] & = & M \int_0^\infty \frac{(1-e^{-\lambda x})^{M-1} \lambda e^{-\lambda x}}{x+b} dx \\
 & = & M \int_0^\infty \frac{e^{-Mt}}{b-\frac{1}{\lambda} \ln(1-e^{-t})} dt
\end{eqnarray*}
where in the second line we used the substitution $e^{-t} = 1-e^{-\lambda x}$. 
To determine the asymptotic behaviour of the integral $$\int_0^\infty \frac{e^{-Mt}}{b-\frac{1}{\lambda} \ln(1-e^{-t})} dt,$$ we will use a Tauberian theorem for the Laplace transform, see p.445 of \cite{Feller2} or p.248 of \cite{Hughes1}, which in our notation says that if $f(t) \geq 0$, $0 \leq \rho < \infty$, and $L(t)$ is a slowly varying function at infinity, then each of the relations 
$$\int_0^\infty e^{-Mt} f(t) dt \sim M^{-\rho} L\left(\frac{1}{M}\right) \textrm{ as } M \rightarrow \infty$$ 
and 
$$\int_0^t f(\tau) d\tau \sim \frac{t^\rho L(t)}{\Gamma(\rho+1)} \textrm{ as } t \rightarrow 0$$
implies the other. 

Thus we can study first the asymptotic behaviour of 
$$\int_0^t \frac{1}{b-\frac{1}{\lambda} \ln(1-e^{-\tau})} d\tau = \lambda \int_0^t \frac{1}{\lambda b- \ln(1-e^{-\tau})} d\tau$$  as $ t \rightarrow 0$.
Using an integration by parts, we obtain
\begin{equation*}
\begin{split}
& \int_0^t \frac{1}{\lambda b-\ln(1-e^{-\tau})} d\tau = \\ &\frac{t}{\lambda b-\ln(1-e^{-t})} - \int_0^t \frac{\tau e^{-\tau}}{(1-e^{-\tau})(\lambda b-\ln(1-e^{-\tau}))^2} d\tau
\end{split}
\end{equation*}
Next, it may be verified  that $\tau e^{-\tau}/(1-e^{-\tau}) \leq 1$, and that $1/(\lambda b-\ln(1-e^{-\tau}))^2$ is an increasing function of $\tau$. Then 
\begin{equation*}
\begin{split}
 &  \left| \int_0^t \frac{\tau e^{-\tau}}{(1-e^{-\tau})(\lambda b-\ln(1-e^{-\tau}))^2} d\tau \right| \\&  \phantom{aaa} =  \int_0^t \frac{\tau e^{-\tau}}{(1-e^{-\tau})(\lambda b-\ln(1-e^{-\tau}))^2} d\tau \\
 & \phantom{aaa} \leq  \int_0^t \frac{1}{(\lambda b-\ln(1-e^{-\tau}))^2} d\tau \\
& \phantom{aaa} \leq   \frac{t}{(\lambda b-\ln(1-e^{-t}))^2} \\
 & \phantom{aaa} =  o\left(\frac{t}{\lambda b-\ln(1-e^{-t})}\right) \textrm{ as } t \rightarrow 0 
\end{split}
\end{equation*}
and so
\begin{equation*}
\begin{split}
\lambda \int_0^t \frac{1}{\lambda b-\ln(1-e^{-\tau})} d\tau & \sim \frac{\lambda t}{\lambda b-\ln(1-e^{-t})} \\ & \sim \frac{\lambda t}{\lambda b-\ln(t)} = \frac{\lambda t}{\lambda b+\ln(1/t)} 
\end{split}
\end{equation*}
as $ t \rightarrow 0$. With $L(t)= \frac{\lambda }{\lambda b+\ln(1/t)}$ (which can be easily shown to be slowly varying) and $\rho=1$, we thus have by the Tauberian theorem that
\begin{equation*}
\begin{split}
M \int_0^\infty \frac{e^{-Mt}}{ b-\frac{1}{\lambda }\ln(1-e^{-t})} dt & \sim M \times M^{-1} \times \frac{\lambda }{\lambda b+\ln(M)} \\ & \sim \frac{\lambda }{\ln(M)} \textrm{ as } M \rightarrow \infty
\end{split}
\end{equation*}

\end{proof}

\subsection{Proof of Lemma \ref{diversity_optimal_lemma}}
\label{diversity_optimal_lemma_appendix}
\begin{proof}
The proof is by contradiction. 
Regard $\nu(M)$ as a function of $M$.

 Suppose $\nu(M)$ does not converge to 0 as $M \rightarrow \infty$. Then there exists a constant $\bar{\nu} > 0$ such that $\nu(M) \geq \bar{\nu}$ for infinitely many values of $M$. In particular, there are infinitely many values of $M$ such that the following is true:
\begin{equation*}
\begin{split}
\int_{b^2 \nu(M)}^\infty \sqrt{\frac{1}{x \nu(M)}} M (1-e^{-\lambda x})^{M-1} \lambda e^{-\lambda x} dx & \leq \frac{1}{\sqrt{\bar{\nu}}} \int_{b^2 \bar{\nu}}^\infty \frac{1}{\sqrt{x}} M (1-e^{-\lambda x})^{M-1} \lambda e^{-\lambda x} dx \\
& < \frac{1}{\sqrt{\bar{\nu}}} \int_{0}^\infty \frac{1}{\sqrt{x}} M (1-e^{-\lambda x})^{M-1} \lambda e^{-\lambda x} dx \\
& \sim \sqrt{\frac{\lambda}{\bar{\nu} \ln(M)}}
\end{split}
\end{equation*}
where the last line comes from (\ref{asympt_integral_1}). Since 
$$ \int_{b^2 \nu(M)}^\infty \frac{b}{x} M (1-e^{-\lambda x})^{M-1} \lambda e^{-\lambda x} dx \geq 0,$$
the condition 
$$\int_{b^2 \nu(M)}^\infty \left( \sqrt{\frac{1}{g_{max} \nu(M)}} - \frac{b}{g_{max}} \right) p(g_{max}) dg_{max} = 1 $$
thus cannot be satisfied for all $M$, which is a contradiction. 
\end{proof}

\subsection{Proof of Lemma \ref{aloha_optimal_lemma}}
\label{aloha_optimal_lemma_appendix}
\begin{proof}
Call $a = \max(T, b^2 \nu) = \max(\frac{1}{\lambda}\ln(M), b^2 \nu)$. Then 
$$\int_a^\infty \left( \sqrt{\frac{1}{g_{i} \nu}} - \frac{b}{g_{i}} \right) \lambda e^{-\lambda g_i} dg_i = \sqrt{\frac{\lambda \pi}{\nu}} \textrm{erfc}(\sqrt{\lambda a}) - b \lambda E_1 (\lambda a) $$
From (\ref{aloha_nu_condition}) we obtain
$$ \sqrt{\frac{\lambda \pi}{\nu}} \textrm{erfc}(\sqrt{\lambda a}) = b \lambda E_1 (\lambda a) + \frac{1}{M}.$$
Now by definition of $a$, we have $ a \geq \frac{1}{\lambda} \ln(M)$, or $ e^{-	\lambda a } \leq \frac{1}{M}$. 
Hence 
$$ \sqrt{\frac{\lambda \pi}{\nu}} \textrm{erfc}(\sqrt{\lambda a}) \geq b \lambda E_1 (\lambda a) + e^{-\lambda a}$$
Also note the inequality $\textrm{erfc}(\sqrt{\lambda a}) \leq e^{-\lambda a}$. Then 
$$
\sqrt{\nu}  \leq \frac{\sqrt{\lambda \pi} \textrm{erfc}(\sqrt{\lambda a})}{b \lambda E_1(\lambda a) + e^{-\lambda a}} \leq \frac{e^{-\lambda a} \sqrt{\lambda \pi}}{b \lambda E_1(\lambda a) + e^{-\lambda a}} = \frac{\sqrt{\lambda \pi}}{1+b\lambda e^{\lambda a} E_1(\lambda a)} \leq \sqrt{\lambda \pi}$$
Thus $\sqrt{\nu} \leq \sqrt{\lambda \pi} \leq  \frac{\sqrt{\ln(M)/\lambda}}{b}$ for $M$ sufficiently large, which then proves Lemma \ref{aloha_optimal_lemma}.

\end{proof}

\end{appendix}

\section*{Acknowledgment}
The authors thank Prof. Barry Hughes of the Department of Mathematics and Statistics, University of Melbourne, for his help on the proof of Lemma \ref{gmax_lemma}. 

%\bibliography{IEEEabrv,sensordiversity}
\bibliography{arxiv}

\begin{thebibliography}{10}
\providecommand{\url}[1]{#1}
\csname url@rmstyle\endcsname
\providecommand{\newblock}{\relax}
\providecommand{\bibinfo}[2]{#2}
\providecommand\BIBentrySTDinterwordspacing{\spaceskip=0pt\relax}
\providecommand\BIBentryALTinterwordstretchfactor{4}
\providecommand\BIBentryALTinterwordspacing{\spaceskip=\fontdimen2\font plus
\BIBentryALTinterwordstretchfactor\fontdimen3\font minus
  \fontdimen4\font\relax}
\providecommand\BIBforeignlanguage[2]{{%
\expandafter\ifx\csname l@#1\endcsname\relax
\typeout{** WARNING: IEEEtran.bst: No hyphenation pattern has been}%
\typeout{** loaded for the language `#1'. Using the pattern for}%
\typeout{** the default language instead.}%
\else
\language=\csname l@#1\endcsname
\fi
#2}}

\bibitem{RibeiroGiannakis1}
A.~Ribeiro and G.~B. Giannakis, ``Bandwidth-constrained distributed estimation
  for wireless sensor networks, {Part I}: Gaussian case,'' \emph{{IEEE} Trans.
  Signal Processing}, vol.~54, no.~3, pp. 1131--1143, Mar. 2006.

\bibitem{SchizasGiannakisLuo}
I.~D. Schizas, G.~B. Giannakis, and Z.-Q. Luo, ``Distributed estimation using
  reduced-dimensionality sensor observations,'' \emph{{IEEE} Trans. Signal
  Processing}, vol.~55, no.~8, pp. 4284--4299, Aug. 2007.

\bibitem{XiaoLuo}
J.-J. Xiao and Z.-Q. Luo, ``Decentralized estimation in an inhomogeneous
  sensing environment,'' \emph{{IEEE} Trans. Inform. Theory}, vol.~51, no.~10,
  pp. 3564--3575, Oct. 2005.

\bibitem{WuHuangLee}
J.-Y. Wu, Q.-Z. Huang, and T.-S. Lee, ``Minimal energy decentralized estimation
  via exploiting the statistical knowledge of sensor noise variance,''
  \emph{{IEEE} Trans. Signal Processing}, vol.~56, no.~5, pp. 2171--2176, May
  2008.

\bibitem{MergenTong}
G.~Mergen and L.~Tong, ``Type based estimation over multiaccess channels,''
  \emph{{IEEE} Trans. Signal Processing}, vol.~54, no.~2, pp. 613--626, Feb.
  2006.

\bibitem{LiuElGamalSayeed}
K.~Liu, H.~{El Gamal}, and A.~M. Sayeed, ``Decentralized inference over
  multiple-access channels,'' \emph{{IEEE} Trans. Signal Processing}, vol.~5,
  no.~7, pp. 3445--3455, July 2007.

\bibitem{GastparVetterli}
M.~Gastpar and M.~Vetterli, ``Source-channel communication in sensor
  networks,'' \emph{Springer Lecture Notes in Computer Science}, vol. 2634, pp.
  162--177, Apr. 2003.

\bibitem{Gastpar_JSAC}
------, ``Power, spatio-temporal bandwidth, and distortion in large sensor
  networks,'' \emph{{IEEE} J. Select. Areas Commun.}, vol.~23, no.~4, pp.
  745--754, Apr. 2005.

\bibitem{Gastpar_optimality}
M.~Gastpar, ``Uncoded transmission is exactly optimal for a simple {Gaussian}
  sensor network,'' \emph{{IEEE} Trans. Inform. Theory}, vol.~54, no.~11, pp.
  5247--5251, Nov. 2008.

\bibitem{Xiao_coherent_TSP}
J.-J. Xiao, S.~Cui, Z.-Q. Luo, and A.~J. Goldsmith, ``Linear coherent
  decentralized estimation,'' \emph{{IEEE} Trans. Signal Processing}, vol.~56,
  no.~2, pp. 757--770, Feb. 2008.

\bibitem{Cui_TSP}
S.~Cui, J.-J. Xiao, A.~Goldsmith, Z.-Q. Luo, and H.~V. Poor, ``Estimation
  diversity and energy efficiency in distributed sensing,'' \emph{{IEEE} Trans.
  Signal Processing}, vol.~55, no.~9, pp. 4683--4695, Sept. 2007.

\bibitem{BahceciKhandani}
I.~Bahceci and A.~K. Khandani, ``Linear estimation of correlated data in
  wireless sensor networks with optimum power allocation and analog
  modulation,'' \emph{{IEEE} Trans. Commun.}, vol.~56, no.~7, pp. 1146--1156,
  July 2008.

\bibitem{FangLi}
J.~Fang and H.~Li, ``Power constrained distributed estimation with correlated
  sensor data,'' \emph{{IEEE} Trans. Signal Processing}, vol.~57, no.~8, pp.
  3292--3297, Aug. 2009.

\bibitem{ThatteMitra}
G.~Thatte and U.~Mitra, ``Sensor selection and power allocation for distributed
  estimation in sensor networks: Beyond the star topology,'' \emph{{IEEE}
  Trans. Signal Processing}, vol.~56, no.~7, pp. 2649--2661, July 2008.

\bibitem{LiDai}
W.~Li and H.~Dai, ``Distributed detection in wireless sensor networks using a
  multiple access channel,'' \emph{{IEEE} Trans. Signal Processing}, vol.~55,
  no.~3, pp. 822--833, Mar. 2007.

\bibitem{KnoppHumblet1}
R.~Knopp and P.~A. Humblet, ``Information capacity and power control in
  single-cell multiuser communications,'' in \emph{Proc. IEEE. Int. Conf.
  Commun.}, Seattle, WA, June 1995, pp. 331--335.

\bibitem{ViswanathTseLaroia}
P.~Viswanath, D.~N.~C. Tse, and R.~Laroia, ``Opportunistic beamforming using
  dumb antennas,'' \emph{IEEE Trans. Inform. Theory}, vol.~48, no.~6, pp.
  1277--1294, June 2002.

\bibitem{QinBerry}
X.~Qin and R.~A. Berry, ``Distributed approaches for exploiting multiuser
  diversity in wireless networks,'' \emph{{IEEE} Trans. Inform. Theory},
  vol.~52, no.~2, pp. 392--413, Feb. 2006.

\bibitem{AdireddyTong}
S.~Adireddy and L.~Tong, ``Exploiting decentralized channel state information
  for random access,'' \emph{{IEEE} Trans. Inform. Theory}, vol.~51, no.~2, pp.
  537--561, Feb. 2005.

\bibitem{Sinopoli}
B.~Sinopoli, L.~Schenato, M.~Franceschetti, K.~Poolla, M.~I. Jordan, and S.~S.
  Sastry, ``Kalman filtering with intermittent observations,'' \emph{{IEEE}
  Trans. Automat. Contr.}, vol.~49, no.~9, pp. 1453--1464, Sept. 2004.

\bibitem{Laneman}
J.~N. Laneman, E.~Martinian, G.~W. Wornell, and J.~G. Apostolopoulos,
  ``Source-channel diversity for parallel channels,'' \emph{{IEEE} Trans.
  Inform. Theory}, vol.~51, no.~10, pp. 3518--3539, Oct. 2005.

\bibitem{Kay_estimation}
S.~M. Kay, \emph{Fundamentals of Statistical Signal Processing: Estimation
  Theory}.\hskip 1em plus 0.5em minus 0.4em\relax New Jersey: Prentice Hall,
  1993.

\bibitem{HongLeiChi}
Y.-W.~P. Hong, K.-U. Lei, and C.-Y. Chi, ``Channel-aware random access control
  for distributed estimation in sensor networks,'' \emph{{IEEE} Trans. Signal
  Processing}, vol.~56, no.~7, pp. 2967--2980, July 2008.

\bibitem{Olver}
F.~W.~J. Olver, \emph{Asymptotics and Special Functions}.\hskip 1em plus 0.5em
  minus 0.4em\relax New York: Academic Press, 1974.

\bibitem{GrimmettStirzaker}
G.~R. Grimmett and D.~R. Stirzaker, \emph{Probability and Random Processes},
  3rd~ed.\hskip 1em plus 0.5em minus 0.4em\relax Oxford, Great Britain: Oxford
  University Press, 2001.

\bibitem{HuMoriczTaylor}
T.-C. Hu, F.~M\'{o}ricz, and R.~L. Taylor, ``Strong laws of large numbers for
  arrays of rowwise independent random variables,'' \emph{Acta Math. Hung.},
  vol.~54, no. 1-2, pp. 153--162, 1989.

\bibitem{BertsekasGallager}
D.~Bertsekas and R.~Gallager, \emph{Data Networks}, 2nd~ed.\hskip 1em plus
  0.5em minus 0.4em\relax USA: Prentice-Hall, 1992.

\bibitem{BergerZhangViswanathan}
T.~Berger, Z.~Zhang, and H.~Viswanathan, ``The {CEO} problem,'' \emph{{IEEE}
  Trans. Inform. Theory}, vol.~42, no.~3, pp. 887--902, May 1996.

\bibitem{LeongDeyEvans_TAES}
A.~S. Leong, S.~Dey, and J.~S. Evans, ``Asymptotics and power allocation for
  state estimation over fading channels,'' \emph{{IEEE} Trans. Aerosp.
  Electron. Syst.}, to appear. Available at http://arxiv.org/abs/0803.3850.

\bibitem{GoldsmithVaraiya}
A.~J. Goldsmith and P.~P. Varaiya, ``Capacity of fading channels with channel
  side information,'' \emph{{IEEE} Trans. Inform. Theory}, vol.~43, no.~6, pp.
  1986--1992, Nov. 1997.

\bibitem{Feller2}
W.~Feller, \emph{An Introduction to Probability Theory and Its Applications:
  Volume {II}}, 2nd~ed.\hskip 1em plus 0.5em minus 0.4em\relax New York: John
  Wiley \& Sons, 1971.

\bibitem{Hughes1}
B.~D. Hughes, \emph{Random Walks and Random Environments: Volume 1}.\hskip 1em
  plus 0.5em minus 0.4em\relax New York: Oxford University Press, 1995.

\end{thebibliography}
\bibliographystyle{IEEEtran}

\end{document}